%% file: HIG-18-024_temp.tex
\pdfoutput=1

\documentclass[11pt,twoside,a4paper,cmspaper,final,collab]{cms-tdr}
\def\svnVersion{57b1305-D}\def\svnDate{2020/09/02}\def\cmsCernNoTag{CERN-EP-2020-061}\def\cmsCernDate{\today}\def\cmsMessage{"Published in the Journal of High Energy Physics as \href{http://dx.doi.org/10.1007/JHEP08(2020)139}{\doi{10.1007/JHEP08(2020)139}.}"}

\begin{document}\cmsNoteHeader{HIG-18-024}

\hyphenation{had-ron-i-za-tion}
\hyphenation{cal-or-i-me-ter}
\hyphenation{de-vices}
\RCS$HeadURL$
\RCS$Id$
\newlength\cmsFigWidth
\setlength\cmsFigWidth{0.4\textwidth}

\newcommand{\PGtmu}{\ensuremath{\PGt_{\PGm}}\xspace}
\newcommand{\PGte}{\ensuremath{\PGt_{\Pe}}\xspace}
\newcommand{\PGth}{\ensuremath{\PGt_{\Ph}}\xspace}

\newcommand{\MMs}{\ensuremath{\PGm\PGm}\xspace}
\newcommand{\MMTT}{\ensuremath{\PGm\PGm\PGt\PGt}\xspace}
\newcommand{\mMM}{\ensuremath{m(\PGm\PGm)}\xspace}
\newcommand{\mTT}{\ensuremath{m(\PGtmu\PGth)}\xspace}
\newcommand{\mMMTT}{\ensuremath{m(\PGm\PGm\PGtmu\PGth)}\xspace}
\newcommand{\DRTT}{\ensuremath{\DR(\PGtmu,\PGth)}\xspace}
\newcommand{\mPH}{\ensuremath{m_{\PH}}\xspace}
\newcommand{\mPGa}{\ensuremath{m_{\Pa}}\xspace}
\newcommand{\mPGm}{\ensuremath{m_{\PGm}}\xspace}
\newcommand{\mPGt}{\ensuremath{m_{\PGt}}\xspace}
\newcommand{\mPqb}{\ensuremath{m_{\PQb}}\xspace}
\newcommand{\sSM}{\ensuremath{\sigma_{\text{SM}}}\xspace}
\newcommand{\sPH}{\ensuremath{\sigma_{\PH}}\xspace}

\newcommand{\mathB}{\ensuremath{\mathcal{B}}}

\cmsNoteHeader{HIG-18-024}
\title{Search for a light pseudoscalar Higgs boson in the boosted $\PGm\PGm\PGt\PGt$ final state in proton-proton collisions at $\sqrt{s}=13\TeV$}

\date{\today}

\abstract{
A search for a light pseudoscalar Higgs boson (\Pa) decaying from the 125\GeV (or a heavier) scalar Higgs boson (\PH) is performed using the 2016 LHC proton-proton collision data at $\sqrt{s}=13\TeV$, corresponding to an integrated luminosity of $35.9\fbinv$, collected by the CMS experiment. The analysis considers gluon fusion and vector boson fusion production of the \PH, followed by the decay $\PH\to\Pa\Pa\to\PGm\PGm\PGt\PGt$, and considers pseudoscalar masses in the range $3.6<\mPGa<21\GeV$. Because of the large mass difference between the \PH and the \Pa bosons and the small masses of the \Pa boson decay products, both the $\PGm\PGm$ and the $\PGt\PGt$ pairs have high Lorentz boost and are collimated. The $\PGt\PGt$ reconstruction efficiency is increased by modifying the standard technique for hadronic \PGt lepton decay reconstruction to account for a nearby muon. No significant signal is observed. Model-independent limits are set at 95\% confidence level, as a function of \mPGa, on the branching fraction (\mathB) for $\PH\to\Pa\Pa\to\PGm\PGm\PGt\PGt$, down to $1.5\,(2.0)\times10^{-4}$ for $\mPH=125\,(300)\GeV$. Model-dependent limits on $\mathB(\PH\to\Pa\Pa)$ are set within the context of two Higgs doublets plus singlet models, with the most stringent results obtained for Type-III models. These results extend current LHC searches for heavier \Pa bosons that decay to resolved lepton pairs and provide the first such bounds for an \PH boson with a mass above 125\GeV.
}

\hypersetup{%
pdfauthor={CMS Collaboration},%
pdftitle={Search for a light pseudoscalar Higgs boson in the boosted mu mu-tau tau final state in proton-proton collisions at sqrt(s) = 13 TeV},%
pdfsubject={CMS},%
pdfkeywords={CMS, physics, Higgs, light Higgs}}

\maketitle

\section{Introduction}
Studies of the
properties of the 125\GeV Higgs boson
can be used to constrain models
that include extended Higgs sectors beyond the standard model (SM)~\cite{
Aad:2013wqa,Khachatryan:2016vau,
Sirunyan:2017exp,Aad:2014aba,Aad:2015zhl}.
Examples include an extension of two Higgs doublets models (2HDM)~\cite{Branco:2011iw}
with a scalar singlet (2HDM+S)~\cite{Curtin:2013fra}, the next-to-minimal supersymmetric SM (NMSSM)~\cite{Ellwanger:2009dp}, and pure
Higgs sector models containing additional Higgs fields~\cite{Curtin:2013fra}.
Especially interesting are models with Higgs boson decay modes that are not detected in the standard channels, which focus on decays to SM particle pairs and invisible decay modes.
A recent study by the CMS Collaboration~\cite{Sirunyan:2018koj} considers models where the Higgs sector contains only doublets and singlets, and the various
couplings are otherwise free to vary with respect to their SM values.  That analysis reports an upper limit of 0.47 on the branching fraction (\mathB) of the Higgs boson to undetected modes
(that is, any mode besides $\PGg\PGg$, $\PZ\PZ$, $\PW\PW$, $\PGt\PGt$, and $\PQb\PQb$)
at 95\% confidence level (\CL), when invisible modes are completely excluded.
This upper limit on undetected modes strengthens as the upper limit on invisible modes weakens.

Given the weak limits on the branching fraction to undetected final states,
it is important to explicitly explore all possibilities
for unseen decay modes.  Among the most prominent
possibilities~\cite{Dermisek:2005ar,Dermisek:2006wr} are decays of the type
$\PH\to{\Pa}{\Pa}$ or $\PH\to{\Ph}{\Ph}$~\cite{Chang:2008cw}, where \PH is a scalar Higgs boson and \Pa (\Ph) is a
lighter pseudoscalar (scalar) Higgs boson.
Such decays are possible
in the SM extensions listed above, and generically have large branching fractions
when kinematically allowed.
However, such decays are not possible in
the $CP$-conserving minimal supersymmetric SM (MSSM)~\cite{Gunion:1989we}.
In what follows, we refer to the light \Pa and \Ph bosons collectively as the \Pa boson.
The Higgs boson observed at 125\GeV can be either the lightest or second-lightest scalar~\cite{Ellwanger:2009dp}.
Given observation of the 125\GeV Higgs boson, more recent theoretical
studies~\cite{King:2012tr,Celis:2013rcs,Grinstein:2013npa,Coleppa:2013dya,Chen:2013rba,Craig:2013hca,Wang:2013sha,Curtin:2013fra,Cao:2013gba,Christensen:2013dra,Cerdeno:2013cz,Chalons:2012qe,Ahriche:2013vqa,Baglio:2014nea,Dumont:2014wha}
consider the possible decays of
this Higgs boson to a pair of lighter Higgs bosons.
In all of these models (aside from the MSSM), it is
possible for the lightest Higgs (pseudo)scalar boson to be much lighter than the
SM-like Higgs boson.  If the light Higgs boson is a scalar then the SM-like Higgs boson
should be identified with the second-lightest scalar of the model.
In the specific case of the NMSSM, a light pseudoscalar boson arises naturally
when model parameters are chosen so that there is either a Peccei--Quinn or $R$ global symmetry of the model~\cite{Ellwanger:2009dp,Dermisek:2005ar,Dermisek:2006wr}.
Either symmetry will be spontaneously broken by the Higgs vacuum expectation values leading to a
massless Nambu--Goldstone boson.
After radiative corrections a nearly massless pseudoscalar, the \Pa, emerges.
Experimental search results are typically presented for four types of 2HDM (and thus 2HDM+S),
differentiated by the couplings of SM fermions to the two doublet fields, $\Phi_1$ and $\Phi_2$,
and by their dependence on the ratio of vacuum expectations for the two Higgs doublets, \tanb.
In particular, the NMSSM corresponds to Type-II 2HDM+S, while for Type-III 2HDM+S
only the charged leptons couple to $\Phi_1$, which yields enhanced rates, especially at large values of \tanb.
We note that in searches performed so far, the event selection and detection efficiencies for
the $\Ph\Ph$ case are essentially the same as for $\Pa\Pa$.
In addition, the branching fractions
for \Ph decays are nearly the same as for \Pa decays.
Finally, the possibility of additional scalar Higgs bosons with masses above 125\GeV
is motivated in generic 2HDM+S~\cite{Curtin:2013fra,Bernon:2015qea}.

Limits from the CERN LEP experiments on the production of a light scalar boson~\cite{Buskulic:1993gi,Acciarri:1996um,Abbiendi:2002qp} are
evaded if the \Ph is singlet-dominated, as required in the limit where the
125\GeV state is SM-like~\cite{Cao:2013gba,Dumont:2014wha,King:2014xwa}.
LEP2 limits on a scalar boson decaying to two light pseudoscalars are obtained for Higgs boson mass (\mPH) less than 107\GeV~\cite{Schael:2010aw}.
Several searches for different scenarios involving light (pseudo)scalar bosons
have been performed by the CERN LHC experiments.
The CMS~\cite{Chatrchyan:2012am} (based on Ref.~\cite{Dermisek:2009fd}) 
and LHCb~\cite{Aaij:2018xpt} Collaborations place limits on
the proton-proton ($\Pp\Pp$) production of a light pseudoscalar decaying to $\PGm\PGm$,
$\sigma(\Pp\Pp\to{\Pa})\mathB({\Pa}\to\PGm\PGm)$,
that significantly constrain the MSSM-like fraction of the NMSSM pseudoscalar state,
especially at large \tanb.
Nonetheless, large $\mathB(\PH\to{\Pa}{\Pa})$ remains possible.
Direct constraints on $\mathB(\PH\to{\Pa}{\Pa})$
are obtained by CMS~\cite{Sirunyan:2018mgs} and ATLAS~\cite{Aaboud:2018fvk} based on the $4\PGm$ final
state and by CMS~\cite{Khachatryan:2017mnf} using the $\PGm\PGm\PGt\PGt$, $4\PGt$, and $\PGm\PGm\cPqb\cPqb$
final states.
Analyses especially relevant for pseudoscalar masses, \mPGa, greater than twice the \PGt lepton mass, \mPGt, are based on the
$\PGm\PGm\PGt\PGt$, $\cPqb\cPqb\PGt\PGt$, $4\PGt$, and $4\cPqb$ final states
and have been performed by the CMS~\cite{Sirunyan:2018mbx,Sirunyan:2018pzn,Sirunyan:2019gou} and ATLAS~\cite{Aad:2015oqa,Aaboud:2018esj,Aaboud:2018iil} Collaborations.

The analysis presented in this paper considers
$\PGm\PGm\PGt\PGt$ final states arising from $\PH\to{\Pa}{\Pa}\to\PGm\PGm\PGt\PGt$, where SM-like production of the \PH boson via the dominant gluon fusion (ggF)
and vector boson fusion (VBF) modes are both included~\cite{deFlorian:2016spz}.
This analysis focuses on the pseudoscalar boson mass range 3.6--21\GeV, complementary to
searches, such as Ref.~\cite{Sirunyan:2018mbx}, that focus on heavier pseudoscalar masses.
For light masses, the large Lorentz boost of the \Pa boson causes its decay products to overlap.  In the $\PGm\PGm$ channel,
the standard CMS muon identification has sensitivity to the topology of boosted muon pairs similar to that for an isolated, nonboosted muon pair.
To reconstruct the collimated \PGt lepton pair,
we have developed a boosted \PGt lepton pair reconstruction technique to target the specific decay
where one \PGt lepton decays to a muon and neutrinos, \PGtmu, while the other decays to one or more hadrons and a neutrino, \PGth, thus: ${\Pa}\to\PGtmu\PGth$.
This technique improves upon the standard CMS \PGt lepton reconstruction that is optimized for isolated, nonboosted \PGt leptons.
The $\PGm\PGm\PGtmu\PGth$ channel has greater detection efficiency than final states with {\cPqb} quarks,
which are difficult to reconstruct at low momentum and significant boost,
and has a larger branching fraction than most models with four-muon final states.
The effectiveness of this improved technique also makes possible for the first time the search
for the decays of a heavier Higgs boson to $\Pa\Pa$ in the $\PGm\PGm\PGt\PGt$ final state at low \mPGa,
with $\mPH=300\GeV$ used as a demonstration.
Such an \PH boson generically has a large branching fraction to any kinematically accessible pair of lighter bosons \cite{Bernon:2014nxa,Bernon:2015qea};
the light bosons are highly boosted and the resulting final-state leptons are similarly collimated.
The search is performed using an unbinned parameterized maximum likelihood fit
of signal and background contributions to the two-dimensional (2D) distribution of the \MMs invariant mass \mMM and
the 4-body visible mass \mMMTT.

This paper is organized as follows. A brief description of the CMS detector is given in Section~\ref{sec:detector}.
Section~\ref{sec:samples} summarizes the data and simulated samples used.
Section~\ref{sec:reco} describes the object identification algorithms, including the modified $\PGtmu\PGth$ reconstruction technique,
while Section~\ref{sec:selection} focusses on the event selection.
The background and signal models of the 2D unbinned fit are described in Section~\ref{sec:bkg} and the
treatment of systematic uncertainties are subsequently discussed in Section~\ref{sec:unc}.
The model-independent results, as well as interpretation in the context of several 2HDM+S types, are presented in Section~\ref{sec:results}.
The paper is summarized in Section~\ref{sec:summary}.

\section{The CMS detector}
\label{sec:detector}

The central feature of the CMS apparatus is a superconducting solenoid
of 6\unit{m} internal diameter, providing a magnetic field of 3.8\unit{T}. 
Within the solenoid volume are a silicon pixel and strip tracker, 
a lead tungstate crystal electromagnetic calorimeter (ECAL), 
and a brass and scintillator hadron calorimeter, 
each composed of a barrel and two endcap sections. 
Forward calorimeters extend the pseudorapidity ($\eta$) coverage 
provided by the barrel and endcap detectors. 
Muons are measured in gas-ionization detectors 
embedded in the steel flux-return yoke outside the solenoid.
Events of interest are selected using a two-tiered trigger system~\cite{Khachatryan:2016bia}. 
The first level (L1), composed of custom hardware processors, uses information from the calorimeters and muon detectors 
to select events at a rate of around 100\unit{kHz} within a time interval of less than 4\mus. 
The second level, known as the high-level trigger (HLT), 
consists of a farm of processors running a version of the full event reconstruction software optimized for fast processing, 
and reduces the event rate to around 1\unit{kHz} before data storage.
A more detailed description of the CMS detector, 
together with a definition of the coordinate system used 
and the relevant kinematic variables, 
can be found in Ref.~\cite{Chatrchyan:2008zzk}. 

\section{Data and simulated samples}
\label{sec:samples}

This search uses a sample of $\Pp\Pp$ collisions at the LHC, collected
in 2016 at $\sqrt{s}=13\TeV$, corresponding to an integrated luminosity of 35.9\fbinv.

The acceptance and reconstruction efficiency of the $\PH\to\Pa\Pa\to\PGm\PGm\PGt\PGt$ processes 
are evaluated using simulated events.
These signal processes are generated with \MGvATNLO version 2.2.2~\cite{mg_amcnlo} at
next-to-leading order (NLO).
The \PYTHIA8.205 program~\cite{Sjostrand:2014zea} is used for parton showering, hadronization, and
the underlying event is simulated with the CUETP8M1 tune~\cite{Khachatryan:2015pea}.
The NNPDF3.0~\cite{Ball:2017nwa} set of parton distribution functions is used.
Samples are generated for $3.6<\mPGa<21\GeV$ for the SM-like \PH boson with $\mPH=125\GeV,$
and for $5<\mPGa<21\GeV$ for a heavier \PH boson with $\mPH=300\GeV$.
The ggF Higgs production process is simulated for each sample
with the obtained signal yields scaled 
to the sum of the expected events from ggF and VBF processes.
The VBF Higgs production process is simulated for a subset of the \PH and \Pa boson mass pairs.
The inclusion of the VBF process increases the expected signal yield by $8\,(19)\%$ for $\mPH=125\,(300)\GeV$.
An acceptance correction arising from a small difference in the analysis
acceptance for ggF and VBF events
of 0.5--3.0\%
is applied as a function of Higgs and pseudoscalar boson masses,
with an uncertainty of 0.5\%.
This correction primarily arises from the differences in transverse momentum \pt spectrum of
the generated \PH and \Pa bosons.  These differences have a negligible
effect on the shapes of the reconstructed pseudoscalar mass
distributions that are used to discriminate signal from background.
The $\PW\PH$, $\PZ\PH$, and $\PQt\PAQt\PH$ Higgs boson production modes
do not significantly increase the sensitivity of this search due to
lower cross sections and reduced acceptance
and are not included.

For all processes, the detector response is simulated using a detailed
description of the CMS detector, based on the \GEANTfour
package~\cite{GEANT}, and the event reconstruction is performed with
the same algorithms used for data.
The simulated samples include additional interactions per bunch crossing (pileup) and
are weighted so that the multiplicity distribution matches the measured one,
with an average of about 23 interactions per bunch crossing.

\section{Event reconstruction}
\label{sec:reco}

Using the information from all CMS subdetectors,
a particle-flow (PF) technique is employed to identify and reconstruct the individual particles
emerging from each collision~\cite{Sirunyan:2017ulk}. The particles are classified into
mutually exclusive categories: charged and neutral hadrons,
photons, muons, and electrons.  Jets and \PGth candidates  are
identified algorithmically using the PF-reconstructed particles as
inputs.  
The missing transverse momentum vector \ptvecmiss  is defined as the projection 
onto the plane perpendicular to the beam axis of the negative vector sum of the momenta 
of all reconstructed PF objects in an event. Its magnitude is referred to as \ptmiss.
The primary $\Pp\Pp$ interaction vertex is defined
as the reconstructed vertex with the largest value of summed physics-object $\pt^2$.
The physics objects considered in the vertex determination
are the objects returned by a jet finding algorithm~\cite{Cacciari:2008gp,Cacciari:2011ma}
applied to all charged tracks associated with the vertex,
plus the corresponding associated \ptmiss,
taken as the negative vector sum of the \pt of those jets.
Finally, additional identification criteria are
applied to the reconstructed muons, electrons, photons, \PGth candidates,
jets, and \ptmiss to reduce the frequency of misidentified
objects.  This section details the reconstruction and
identification of muons, jets, and \PGth candidates.

\subsection{Muons}
\label{sec:reco-muons}

Muons are reconstructed within $\abs{\eta(\PGm)}<2.4$~\cite{Sirunyan:2018fpa}.
The reconstruction combines the information from both the tracker and the
muon spectrometer.
The muons are selected from among the reconstructed muon track candidates
by applying minimal requirements on the track components in the muon system
and taking into account matching with small energy deposits in the
calorimeters.  For each muon track,
the distance of closest approach to the primary vertex in the transverse
plane is required to be less than $0.2\cm$.
The distance of closest approach to the primary vertex along the beamline, $d_{\text{z}}$, must be less than $0.5\cm$.

The isolation of individual muons is defined relative to their transverse momentum $\pt(\PGm)$
by summing over the \pt of charged hadrons and neutral particles within a cone
around the muon direction at the interaction vertex 
with radius $\DR=\sqrt{\smash[b]{(\Delta\eta)^2+(\Delta\phi)^2}}<0.4$
(where $\phi$ is the azimuthal angle in radians) :
\begin{equation}
I^{\PGm} = \left( \sum  \pt^\text{charged} + \text{max}\left[ 0, \sum \pt^\text{neutral}
                                 +  \sum \pt^{\gamma} - \pt^{\text{PU}}  \right] \right) /  \pt(\PGm).
\end{equation}
Here, $\sum \pt^\text{charged}$ is the scalar \pt sum of charged hadrons
originating from the primary vertex.
The $\sum \pt^\text{neutral}$ and $\sum \pt^{\gamma}$ are the
scalar \pt sums for neutral hadrons and photons, respectively.
The neutral contribution to the isolation from pileup interactions,
$\pt^{\text{PU}}$, is estimated as
$0.5\sum_i\pt^{\text{PU},i}$,
where $i$ runs over the charged hadrons originating from pileup vertices and
the factor 0.5 corrects for the ratio of charged to neutral particle contributions in the isolation cone.
Muons are considered isolated if $I^{\PGm}<0.25$.

\subsection{Jets}
\label{sec:reco-jets}

Jets are reconstructed using PF objects.
The anti-\kt jet clustering
algorithm~\cite{Cacciari:2008gp, Cacciari:2011ma} with a distance parameter of 0.4 is used. The standard method for jet energy
corrections~\cite{Chatrchyan:2011ds} is applied.
In order to reject
jets coming from pileup collisions, a multivariate (MVA) jet identification
algorithm~\cite{CMS-PAS-JME-18-001} is applied.
This algorithm takes advantage of differences in the shapes of energy
deposits in a jet cone between pileup jets and jets originating from a quark or gluon.
The combined secondary vertices (CSV) {\cPqb} tagging algorithm~\cite{Chatrchyan:2012jua} 
is used to identify jets originating from {\cPqb} hadrons~\cite{Sirunyan:2017ezt}. The efficiency for tagging {\cPqb} jets is ${\approx}63\%$,
while the misidentification probability for charm (light-quark or gluon) jets is ${\approx}12\,(1)\%$.

\subsection{\texorpdfstring{Hadronic $\tau$ lepton decays}{Hadronic tau lepton decays}}
\label{sec:reco-tau}

Hadronically decaying \PGt leptons are reconstructed and identified
within $\abs{\eta(\PGth)}<2.3$ using the hadron-plus-strips (HPS) algorithm~\cite{Sirunyan:2018pgf},
which targets the main decay modes by selecting PF objects with
one charged hadron and up to two neutral pions, or with three charged hadrons.  
The HPS algorithm is seeded by the jets described in
Section~\ref{sec:reco-jets}.  
The \PGth candidates are reconstructed based on the number of tracks and on the
number of ECAL strips with an energy deposit in the $\eta$-$\phi$ plane.

This analysis uses a specialized $\PGtmu\PGth$ reconstruction algorithm,
which uses the same HPS method as the above,
with a modified jet seed.
This method is designed to reconstruct boosted $\PGtmu\PGth$ objects, for which
the \PGt lepton decaying leptonically to a muon overlaps
with the hadronic decay products of the other \PGt lepton.  
One \PGt lepton is required to decay to a muon because this mode has a high reconstruction efficiency
and a low misidentification probability.
As in Ref.~\cite{Khachatryan:2017mnf}, a joint reconstruction
of the \PGth candidate and a nearby muon is performed.
Jets that seed the \PGth reconstruction are first modified to remove
muons with $\pt>3\GeV$ passing minimal identification requirements from their jet constituents.
The \PGth candidates reconstructed using these modified jets are 
required to have $\pt>10\GeV$, where the reconstructed $\pt(\PGth)$
corresponds to the visible portion of the \PGt lepton decay.
To reject \PGth candidates that arise from
constituents not originating from the primary vertex,
the \PGth candidates must have $d_{\mathrm{z}}<0.5\cm$.
To reduce background contribution from jets arising from {\cPqb} quarks, 
the jet seeds to the \PGth reconstruction must additionally fail
the CSV jet tagging algorithm.
Because no MVA discriminant to reject electrons~\cite{Sirunyan:2018pgf} is applied,
the \PGth reconstruction algorithm has high efficiency to select
\PGt leptons that decay to electrons, \PGte.
The fraction of reconstructed \PGth candidates that are 
\PGte decays is estimated from simulation to be 18--22\%, 
predominantly reconstructed in the one-prong decay mode with no additional neutral hadrons.
No distinction is made between \PGte and \PGth candidates 
and this paper refers to the contribution of both decay categories
as \PGth candidates.

The full $\PGtmu\PGth$ identification procedure includes the modified HPS
algorithm described above, along with a requirement on the \PGth candidate
isolation.  The isolation of a \PGth candidate is computed using 
an MVA discriminant~\cite{Sirunyan:2018pgf}.
The discriminant is computed using PF candidates,
with the overlapping muon excluded,
in the region around the \PGth candidate
defined by $\DR<0.8$.  The \PGth candidates are required to pass
a selection on the MVA discriminant output as a function of $\pt(\PGth)$ to yield an
approximately constant efficiency of ${\approx}80\%$.
No discriminant to reject muons~\cite{Sirunyan:2018pgf} is applied,
as it would reduce the reconstruction efficiency of the
boosted $\PGtmu\PGth$ final state.

\subsection{Charged lepton efficiency}
\label{sec:reco-SF}

The combined efficiencies of the reconstruction, identification, and isolation
requirements for muons are measured
in several bins of $\pt(\PGm)$ and $\abs{\eta(\PGm)}$ using a ``tag-and-probe''
technique~\cite{CMS:2011aa} applied to an inclusive sample of muon pairs from \PZ boson and \PJGy meson events~\cite{Sirunyan:2018fpa}.
These efficiencies are measured in data and simulation.
The data to simulation efficiency ratios are used as scale factors
to correct the simulated event yields.
For \PGth candidates, two scale factors are similarly measured
using a $\PZ\to\PGtmu\PGth$ sample~\cite{Sirunyan:2018pgf}
to be $0.60\pm0.11\,(0.97\pm0.05)$ for $10<\pt(\PGth)<20\GeV\,(\pt(\PGth)>20\GeV)$,
which are found to be independent of $\abs{\eta(\PGth)}$.
For $10<\pt(\PGth)<20\GeV$, the
$\PZ\to\PGtmu\PGth$ data sample contains significant
$\PW$+jets background, making the scale factor difficult to estimate
with as high a precision as for $\pt(\PGth)>20\GeV$.

\section{Event selection}
\label{sec:selection}

Collision events are selected by a trigger that requires
the presence of an isolated muon with
$\pt>24\GeV$~\cite{Khachatryan:2016bia}.
Trigger efficiencies are measured in data and simulation
using the tag-and-probe technique.
The event is required to have two isolated opposite-sign muons with $\DR<1$.
The leading muon which is matched to the muon that triggered the event
must have $\pt>26\GeV$.
The second muon must have $\pt>3\GeV$.
These muons constitute a $\PGm\PGm$ pair from
one of the pseudoscalar candidates.

The second pseudoscalar is selected via its decay to an isolated opposite-sign $\PGtmu\PGth$ pair.  
The $\PGtmu\PGth$ selection requires one identified muon with $\pt>3\GeV$, with no isolation
selection imposed, and one \PGth candidate with $\pt>10\GeV$,
reconstructed as described in Section~\ref{sec:reco-tau}.  The
reconstructed muon corresponds to the visible
portion of the \PGtmu decay.  
The two \PGt lepton candidates are required to lie within $\DRTT<0.8$.
The value of $0.8$ is driven by the modified HPS algorithm isolation discriminant
and ensures the boosted topology.
This selection, with the corresponding selection of the $\PGm\PGm$ pair,
prevents combinatoric background in which the wrong combination of leptons
is assigned to the pseudoscalar candidates.
The $\PGm\PGm$ pair selection is looser to avoid loss of efficiency.

\begin{figure}
  \centering
  \includegraphics[scale=0.5]{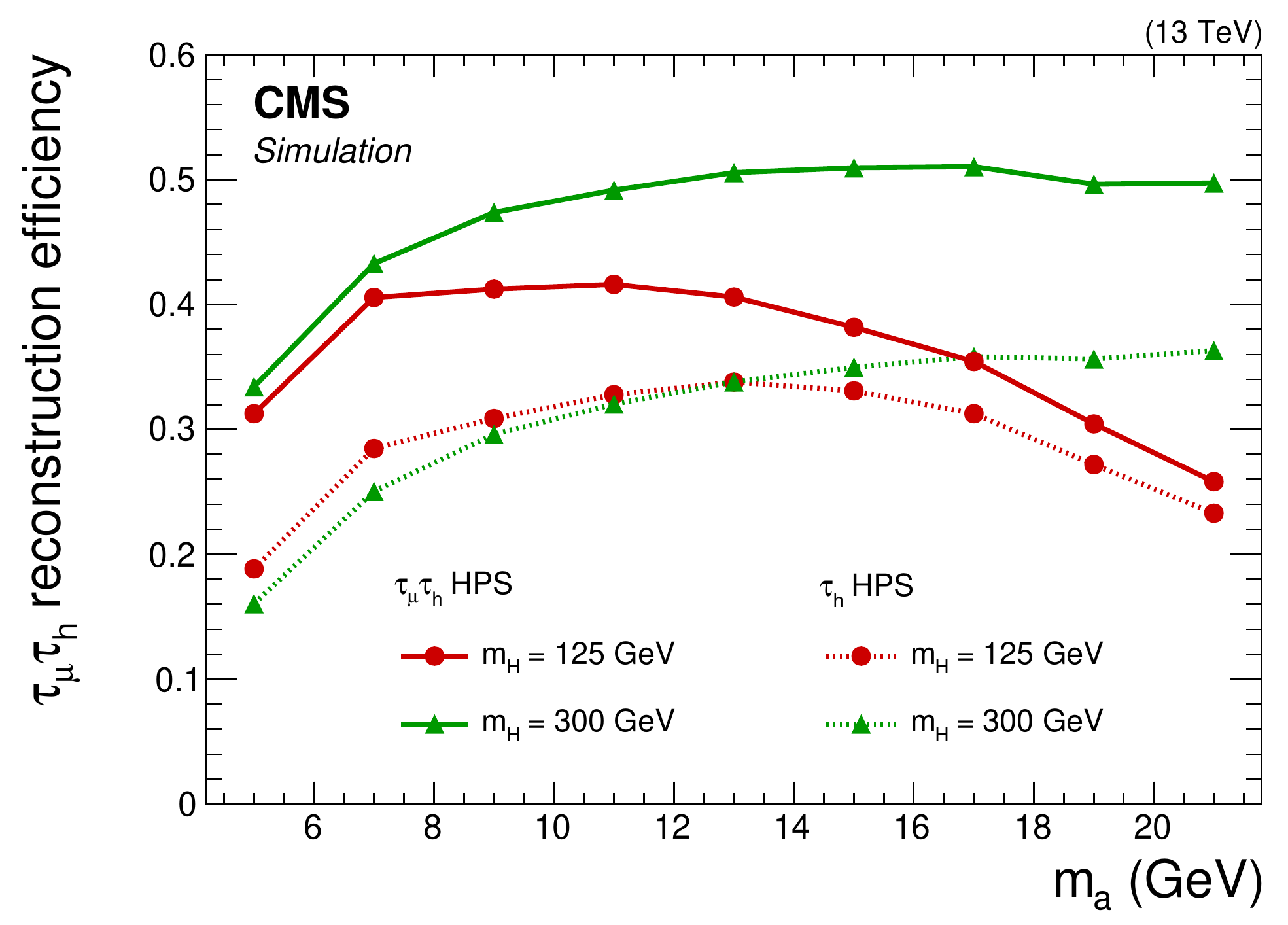}
  \caption{
          The efficiency of the standard HPS (dashed lines) and $\PGtmu\PGth$ HPS reconstruction used in this search (solid lines)
          as a function of pseudoscalar boson mass for $\mPH=125$~(red) and 300\GeV~(green).
          The events are required to have two reconstructed muons passing identification and isolation
          criteria. The efficiency is measured by additionally requiring a third muon passing identification
          requirements and a \PGth candidate reconstructed using either the standard HPS algorithm or
          the $\PGtmu\PGth$ HPS algorithm and passing isolation requirements.
          }
  \label{fig:CleanedEff}
\end{figure}

The modified $\PGtmu\PGth$ reconstruction and identification algorithm 
increases the signal efficiency throughout the full
range of Higgs boson and pseudoscalar hypotheses considered, as shown in Fig.~\ref{fig:CleanedEff}.
The efficiency of the $\PGtmu\PGth$ reconstruction and identification
is measured by requiring the
presence of a muon passing the identification requirements
and a \PGth candidate passing
either the standard \PGth HPS reconstruction
or the $\PGtmu\PGth$ HPS reconstruction,
as well as the MVA isolation discriminant.
The increase in efficiency arises incrementally
both from the modification of the jets which seed
the $\PGtmu\PGth$ reconstruction and the exclusion
of the muon energy from the MVA isolation discriminant.
Because of the increase in Lorentz boost,
the jet seed modification
is the primary cause of increased efficiency
at low \mPGa where the pseudoscalar decay products are most overlapping,
with $\DRTT<0.4$.
At larger separation, $0.4<\DRTT<0.8$, the change in the MVA discriminant
becomes the only source of efficiency increase.
The reduced efficiency
at low pseudoscalar mass is due to the high Lorentz boost
in which the muon is nearly collinear with a charged hadron
from the \PGth candidate.
At low Lorentz boost, the muon and \PGth candidate
have a large separation.
In this case, the efficiency is reduced from the 
requirement of the boosted topology,
especially at $\mPH=125\GeV$.  
The efficiency for the higher \PH boson mass is less
affected by an increase in pseudoscalar mass because the reduction in
Lorentz boost is generally not significant enough to separate the \PGt leptons from
a pseudoscalar decay beyond the selection requirement of $\DRTT<0.8$.

\section{Signal and background modeling}
\label{sec:bkg}

The main source of background in this search is Drell--Yan \MMs
production in association with at least one jet that is misidentified as the
$\PGtmu\PGth$ candidate.  This background, reduced by the $\PGtmu\PGth$ reconstruction,
features the prominent
\MMs resonances with masses between 3.6 and 21\GeV:
$\PGyP{2S}\,(3.69\GeV)$, $\PGUP{1S}\,(9.46\GeV)$,
$\PGUP{2S}\,(10.0\GeV)$, and $\PGUP{3S}\,(10.4\GeV)$~\cite{PDG2018}.
In the \mMM distribution, the known resonance peaks
appear on top of the Drell--Yan continuum.  In the \mMMTT distribution,
the $\MMs+\text{jet}$ background
appears as an exponentially falling distribution with a threshold around
40--60\GeV because of the \pt
thresholds of the three reconstructed muons and one
\PGth candidate.  The signal is characterized by a narrow \mMM resonance
from a pseudoscalar decay and a broader \mMMTT distribution
because of the invisible decay products of one of the pseudoscalar Higgs bosons.
As described below,
the search strategy consists of an unbinned fit of \mMM vs. \mMMTT,
using analytical models for the signal and
background shapes in each dimension.  
The background shape model for the Drell--Yan
continuum, the meson resonances mentioned above, and additionally the
\PJGy resonance (3.10\GeV~\cite{PDG2018}) are constrained via a
data control region enriched in $\PGm\PGm$+jet events.
Although the
\PJGy resonance falls outside the kinematically allowed search
window for a $\PGt\PGt$ resonance, it is modeled in the fit to provide a
better background description near the \PGyP{2S} meson.

\begin{figure}[hb]
    \centering

    \includegraphics{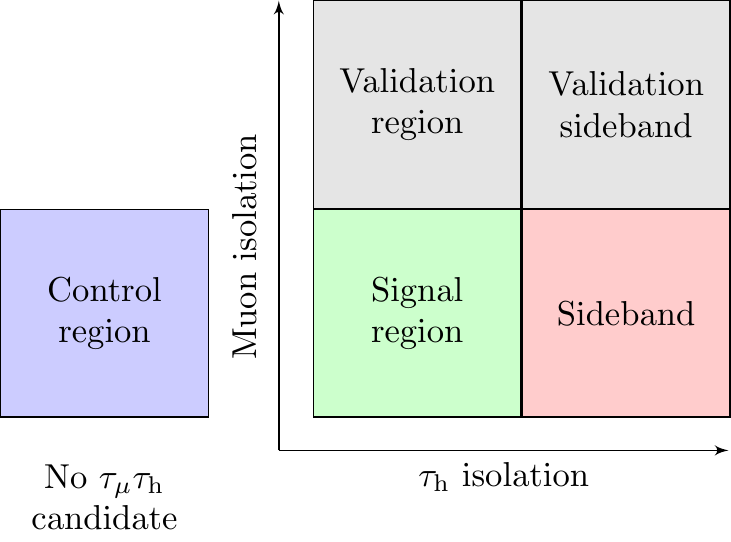}

    \caption{
    Schematic of the fit regions in the analysis.
    Events with two isolated muons and no $\PGtmu\PGth$ candidates constitute the control region (blue).
    Events that have a $\PGtmu\PGth$ candidate are further divided based on the isolation of the \PGth candidate
    with isolated $\PGtmu\PGth$ candidates forming the signal region (green) and the remaining $\PGtmu\PGth$ candidates forming the sideband (red).
    Additionally, the $\PGm\PGm$ candidates that fail the muon isolation selection form two
    analogous regions for the validation of the background fit model (gray).
    }
    \label{fig:regions}
\end{figure}

The analysis uses a simultaneous unbinned fit of three mutually exclusive regions to 
model the background and search for a signal.
The ``control region'' requires the presence of two muons
and no identified $\PGtmu\PGth$ candidate.
The next two regions additionally require a reconstructed
$\PGtmu\PGth$ candidate and are defined by passing or failing
the \PGth MVA isolation requirement,
labeled as ``signal region'' and ``sideband'', respectively.
A schematic depiction of the three regions is shown in Fig.~\ref{fig:regions}.
Two additional regions are also shown and are used
to validate the background estimation method described below.

The choice of \mMM and \mMMTT
as observables for distinguishing the $\PH\to\Pa\Pa$ signal
from the SM background processes
is found to be more performant
than combinations including \mTT
over the largest range of Higgs boson
and pseudoscalar mass hypotheses.
The signal is modeled as a 2D function given by the product of a Voigt function for \mMM
and a split normal distribution for \mMMTT.
For the signal processes, there is minimal correlation between
the \mMM and \mMMTT distributions.
The parameters of the model are
determined from fits to the signal simulation.
Each generated distribution, with a specified Higgs boson and pseudoscalar mass,
is fit with the described 2D function.
For each parameter, a polynomial function is used to interpolate
between the generated masses:
a first-order polynomial for the mean value of the \mMM and \mMMTT,
a second-order polynomial for each width parameter,
and the product of a first-order polynomial and two error functions for the signal normalization.
The search is performed for pseudoscalar masses between 3.6 and $21\GeV$.

The 2D fit of \mMM vs. \mMMTT
is performed in data to model the SM background processes
and extract any significant signal process contribution
in three ranges of the \mMM spectrum: 
$2.5<\mMM<8.5\GeV$, $6<\mMM<14\GeV$,
and $11<\mMM<25\GeV$.
For a given \mPGa,
a single \mMM range is used, with
the transition between the \mMM ranges
occurring at $\mPGa=8$ and $11.5\GeV$.
There is some overlap in the fit ranges
to allow the lower or upper portion of the signal
model to be fully contained in the given fit range.
The background probability density function (PDF) used
for the \mMM spectrum is the sum of an exponential together with
two, three, or zero Voigt distributions to model the SM resonances for the three respective ranges.
An additional exponential function is necessary
to model the rising continuum background near the \PJGy resonance
in the lowest \mMM range.
The \mMMTT background distribution
is modeled with the product of an error function and the sum of two exponential
distributions. 
The second exponential provides the fit with additional flexibility
to allow the fit to favor an extended tail if necessary.
The fit range is $0<\mMMTT<1200\GeV$
in all three \mMM ranges.
The \mMM and \mMMTT functions are multiplied together to produce a 2D PDF.
Because \mMMTT is loosely correlated
with \mMM in the background distribution,
the parameters of the \mMMTT background model in
a given \mMM range are allowed to vary
independently of the other ranges, allowing
a correlation between \mMM and \mMMTT.

The normalization of the background
model in the signal region is estimated from
the sideband using a ``tight-to-loose'' method.
This method uses a $\PZ(\MMs)+\text{jet}$
sample to estimate the efficiency for a jet
that has passed all the \PGth reconstruction requirements (including the muon removal step) of
Section~\ref{sec:reco-tau}, except the MVA isolation requirement,
to additionally pass the MVA isolation requirement.
The region contains events collected with a single muon
trigger with the requirement of two isolated 
opposite-sign muons and a jet that has been misidentified
as a $\PGtmu\PGth$ object with a muon within $\DRTT<0.8$,
without the requirement on the MVA isolation.
The $\PGm\PGm$ pair must have invariant mass $81<\mMM<101\GeV$.
The tight-to-loose ratio, $f$,
is defined as the ratio of the number of \PGth candidates
that pass the MVA isolation
requirement in addition to the other identification
requirements (the ``tight'' condition) to the number of
\PGth candidates
that pass the other identification requirements, but with a relaxed requirement on the isolation
(the ``loose'' condition).
The calculation of $f$ is performed separately
for each hadronic decay mode of the \PGt lepton
and is binned in $\pt(\PGth)$.
This region is dominated by Drell--Yan events containing jets.
Residual contributions from diboson processes,
as estimated from simulation, are subtracted from the data.
The associated jets are the objects most likely to pass the \PGth reconstruction criteria.
This tight-to-loose ratio
is measured to be 10--40\%,
increasing at lower $\pt(\PGth)$.
In general, the decay mode with three charged tracks
has a lower tight-to-loose ratio than
those with a single charged track.

The sideband is then reweighted using the tight-to-loose method
to estimate the contribution in the signal region.
The weights are applied on an event-by-event basis as a function of $\pt(\PGth)$.
The tight-to-loose method
is verified in a validation region independent of
the analysis region by inverting
the isolation requirement on the
muon in the \MMs pair that did not trigger the event.
These regions correspond to
the gray boxes in Fig.~\ref{fig:regions}.
The expected and observed yields in
this validation region are compatible within 15\%,
and an uncertainty is derived from this value.

The parameters of the \MMs resonances---mean ($\mu$), width ($\Gamma$), and resolution ($\sigma$)---and
the relative normalizations---$N_{i}/N_{j}$ where $i$ and $j$ are a pair of background
resonances---between the \PJGy and \PGyP{2S} resonances
and between the \PGUP{1S} and each of the \PGUP{2S} and \PGUP{3S} resonances
are constrained via a simultaneous fit among all three regions.
The parameters of the resonances 
are compatible, and thus the same, among the three regions,
while their relative normalizations are only the same in the sideband
and control region with the signal region relative normalizations
related to the sideband via a linear transformation.
The slope and constant values of this
linear transformation are
determined from a fit to the sideband and the tight-to-loose estimation
of the background in the signal region.
An uncertainty is assigned for
this linear constraint in the signal region.
This uncertainty is derived in a validation region and a corresponding validation sideband in which
the muon of the \MMs pair which did not trigger the event has an inverted isolation requirement
and is measured to be 5--20\% depending on the resonance.
The parameters of the \MMs continuum ($\lambda^{i}_{\MMs}$), the
\mMMTT continuum ($\lambda^{i}_{\MMTT}$), 
the \mMMTT  error function shift ($\erf_a$) and scale ($\erf_b$),
and the relative normalizations of the \MMs resonances to the \MMs continuum ($N_{\PGUP{1S}}/N_{\PJGy}$ and  $N_{\PJGy}/N_{\text{continuum}}$)
are constrained in the signal region to the sideband via the tight-to-loose method.
All remaining parameters are free to vary independently of each other and share no constraint between regions.
Table~\ref{tab:parameters} summarizes these constraints.

\begin{table}[!htp]
    \centering
    \topcaption{
    Background model parameters and their relations among the three fit regions in the analysis.
    The \MMs background model includes the five meson resonances modeled using a Voigt function over an exponential continuum.
    The 4-body background model includes an error function multiplied with the sum of two exponential distributions.
    Three types of fit region relations are used: (a) constrained, in which the parameters are the same in the indicated regions,
    (b) free, in which the parameter is not related to those in any other
    region, and (c) related via the $\PGtmu\PGth$ tight-to-loose ratio,
    in which the indicated parameter in the signal region is constrained to the corresponding parameter
    in the sideband via a linear transformation.
    \label{tab:parameters}}

    \begin{tabular}{ccccc}
    \hline

    Category & Parameters & Signal region & Sideband & Control region \\ \hline

    $\PGm\PGm$ resonances & $\mu$, $\sigma$, $\Gamma$  & \multicolumn{3}{c}{Constrained (three regions)} \\
    $\PGm\PGm$ continuum  & $\lambda^{\mathrm{i}}_{\MMs}$                 & Tight-to-loose & Free & Free \\
    $\PGm\PGm\PGtmu\PGth$  & $\mathrm{Erf}_a$, $\mathrm{Erf}_b$, $\lambda^{\mathrm{i}}_{\MMTT}$      & Tight-to-loose & Free & \NA \\
    Normalizations        & $N_{\PGyP{2S}}/N_{\PJGy}$           & Tight-to-loose & \multicolumn{2}{c}{Constrained (two regions)} \\
                          & $N_{\PGUP{2S}}/N_{\PGUP{1S}}$        & Tight-to-loose & \multicolumn{2}{c}{Constrained (two regions)} \\
                          & $N_{\PGUP{3S}}/N_{\PGUP{1S}}$      & Tight-to-loose & \multicolumn{2}{c}{Constrained (two regions)} \\
                          & $N_{\PGUP{1S}}/N_{\PJGy}$           & Tight-to-loose & Free & Free \\
                          & $N_{\PJGy}/N_{\text{continuum}}$    & Tight-to-loose & Free & Free \\ 

    \hline

    \end{tabular}
\end{table}

\begin{figure}[htbp]
  \centering
  \includegraphics[width=0.4\textwidth]{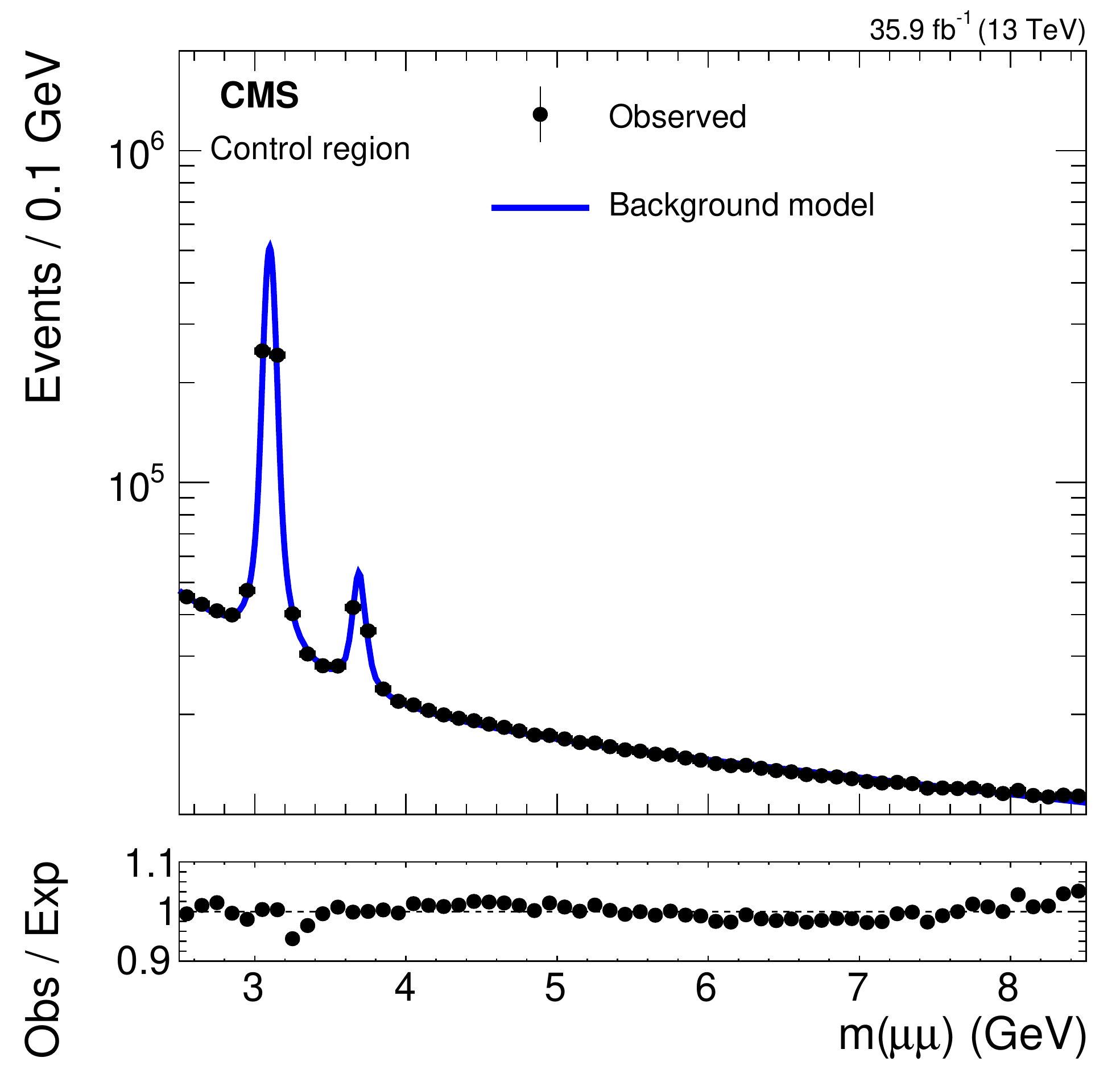}
  \includegraphics[width=0.4\textwidth]{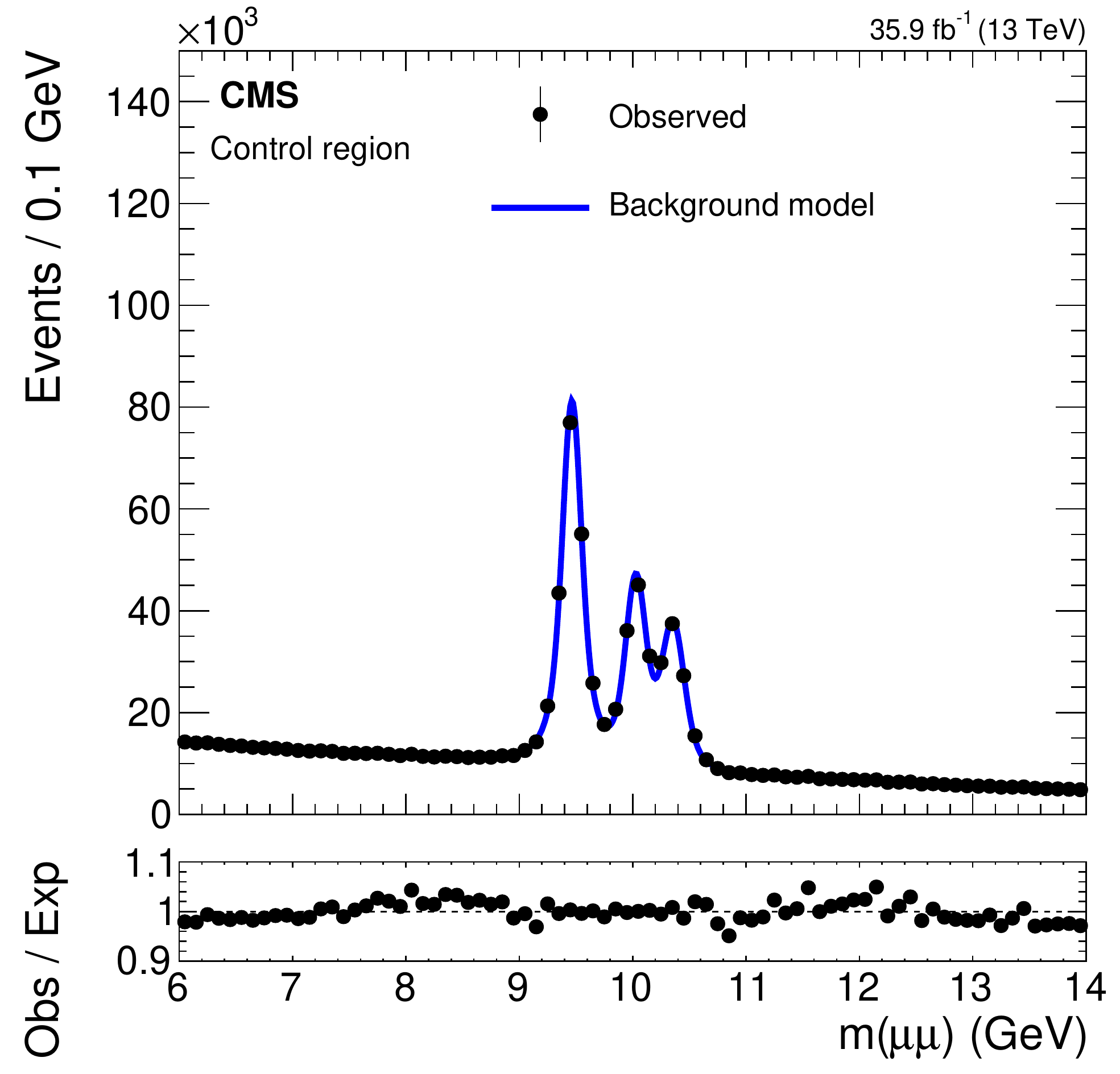} \\
  \includegraphics[width=0.4\textwidth]{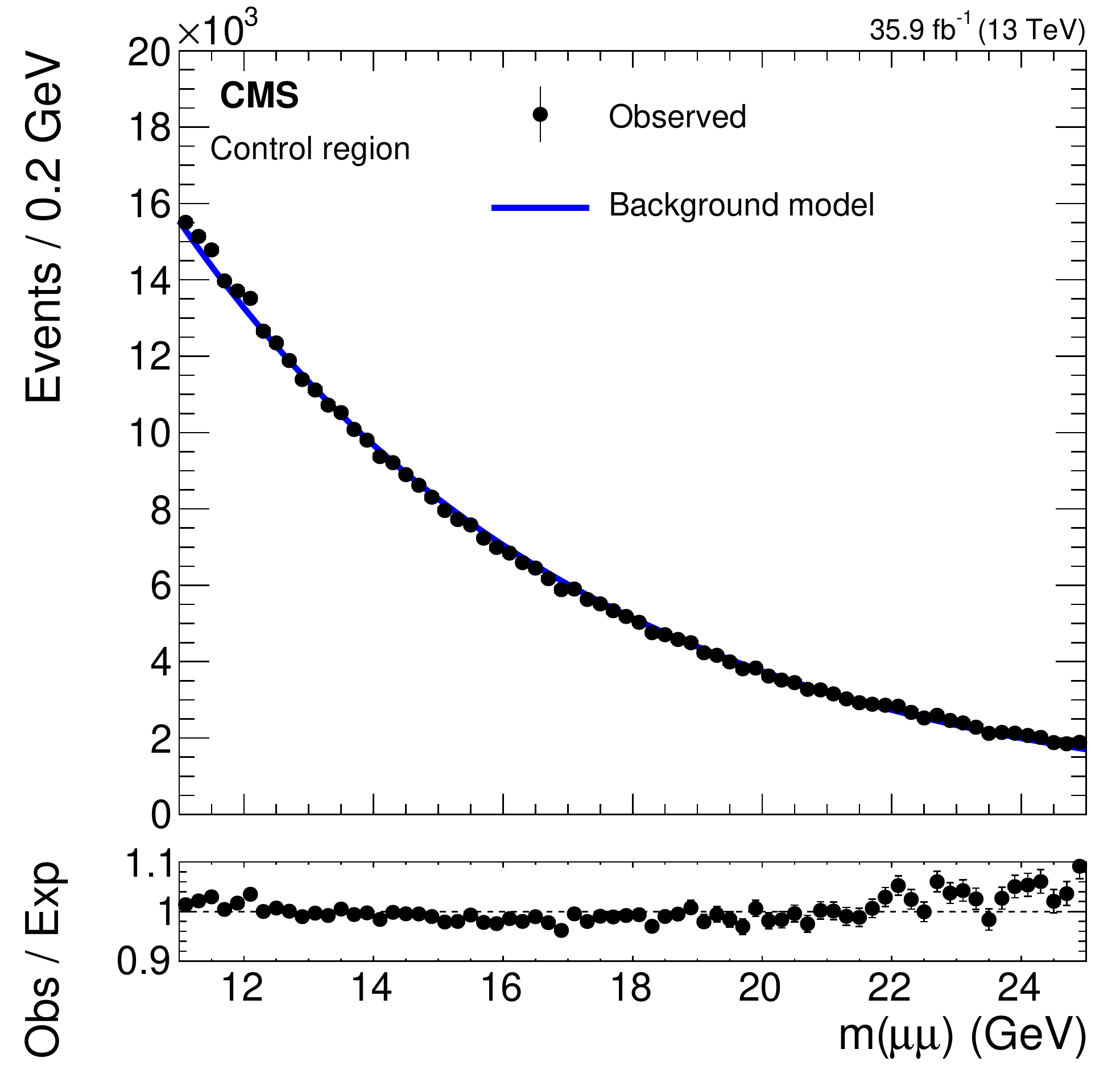}
  \caption{
  Background model fits and observed data in the control region \mMM distribution.
  The figures are divided into three fit ranges:
  $2.5<\mMM<8.5\GeV$ (upper-left), $6<\mMM<14\GeV$ (upper-right),
  and $11<\mMM<25\GeV$ (lower).
           }
  \label{fig:bgfitscontrol}
\end{figure}

\begin{figure}[htbp]
  \centering
  \includegraphics[width=0.4\textwidth]{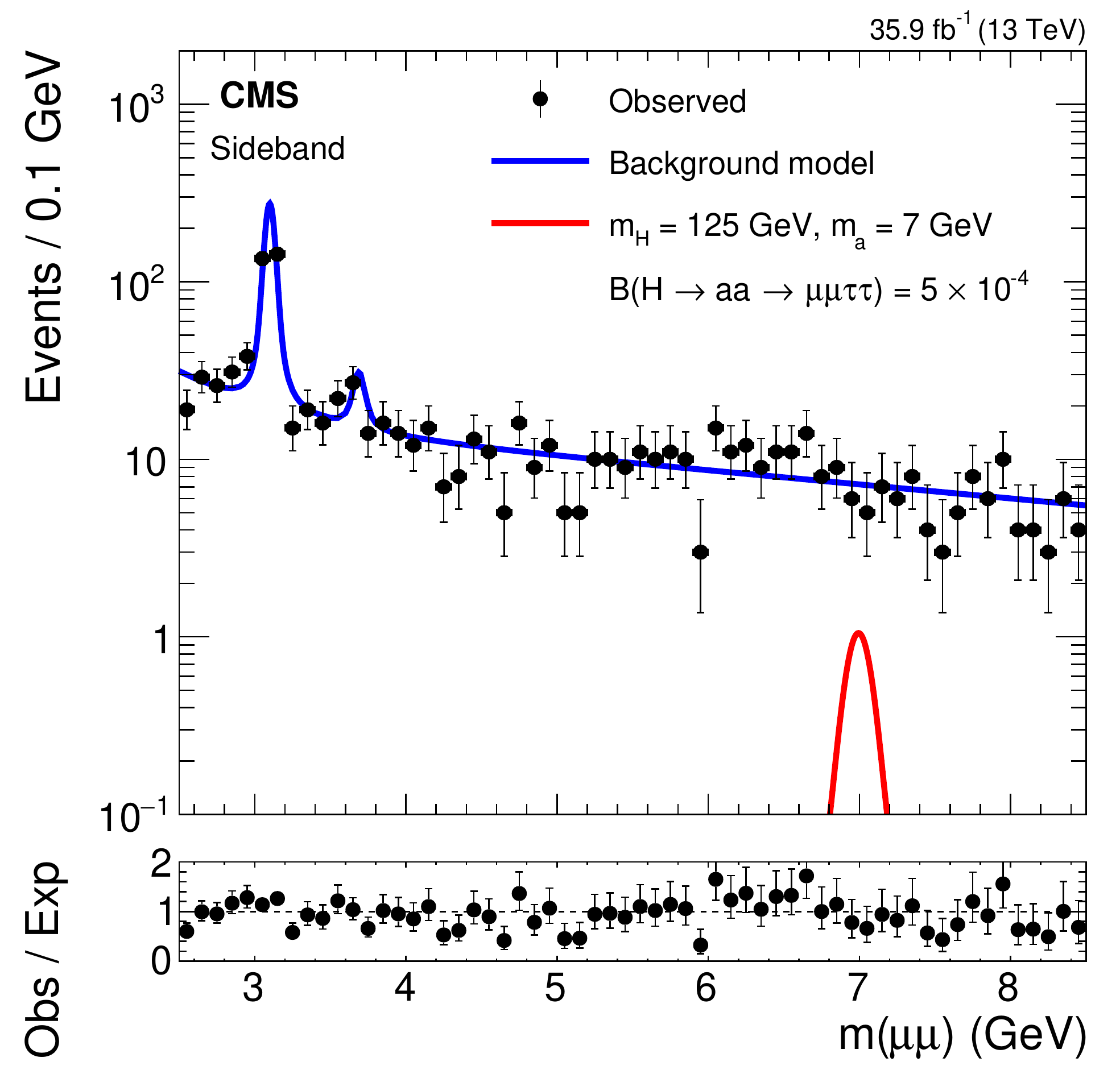}
  \includegraphics[width=0.4\textwidth]{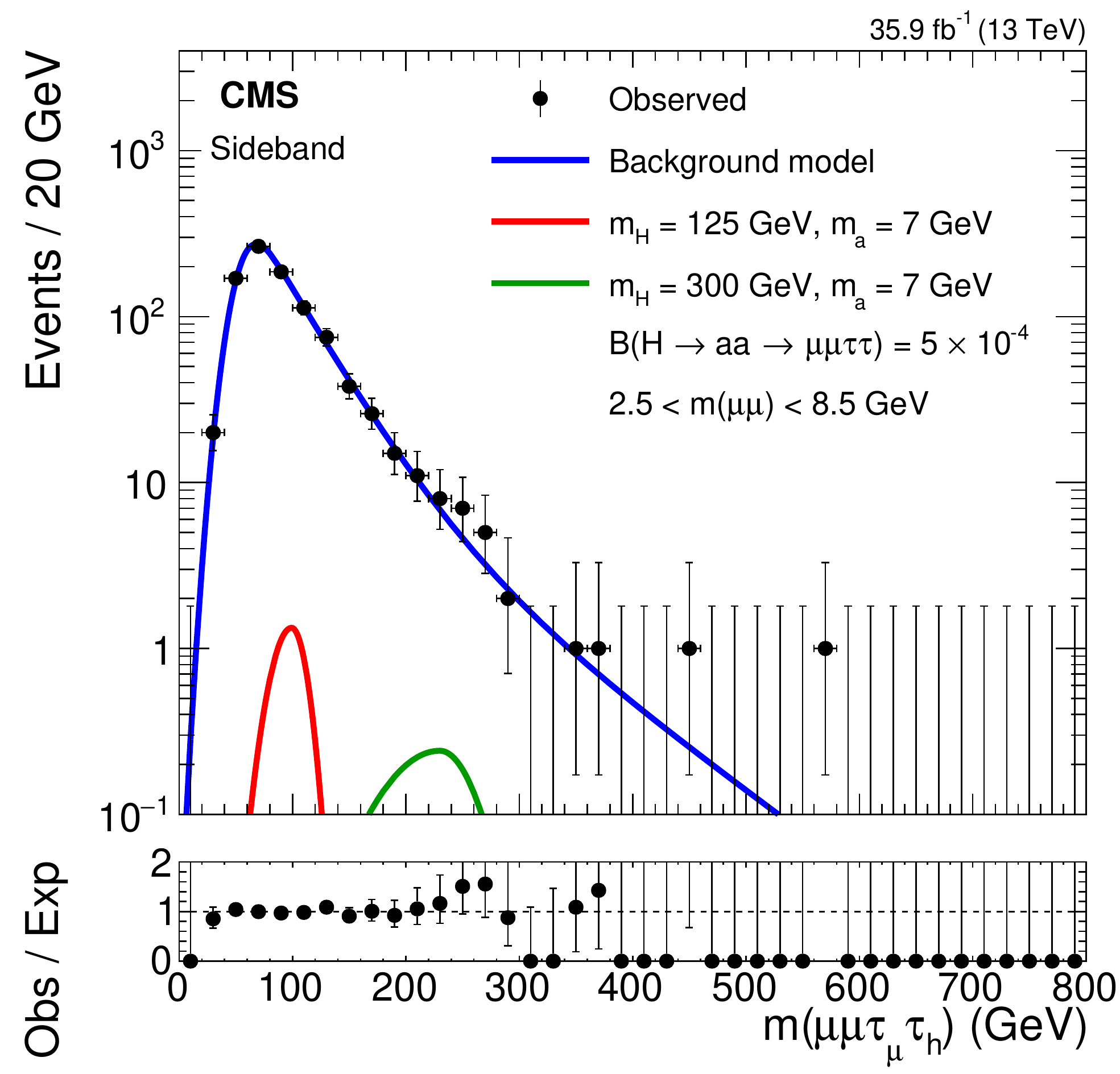} \\
  \includegraphics[width=0.4\textwidth]{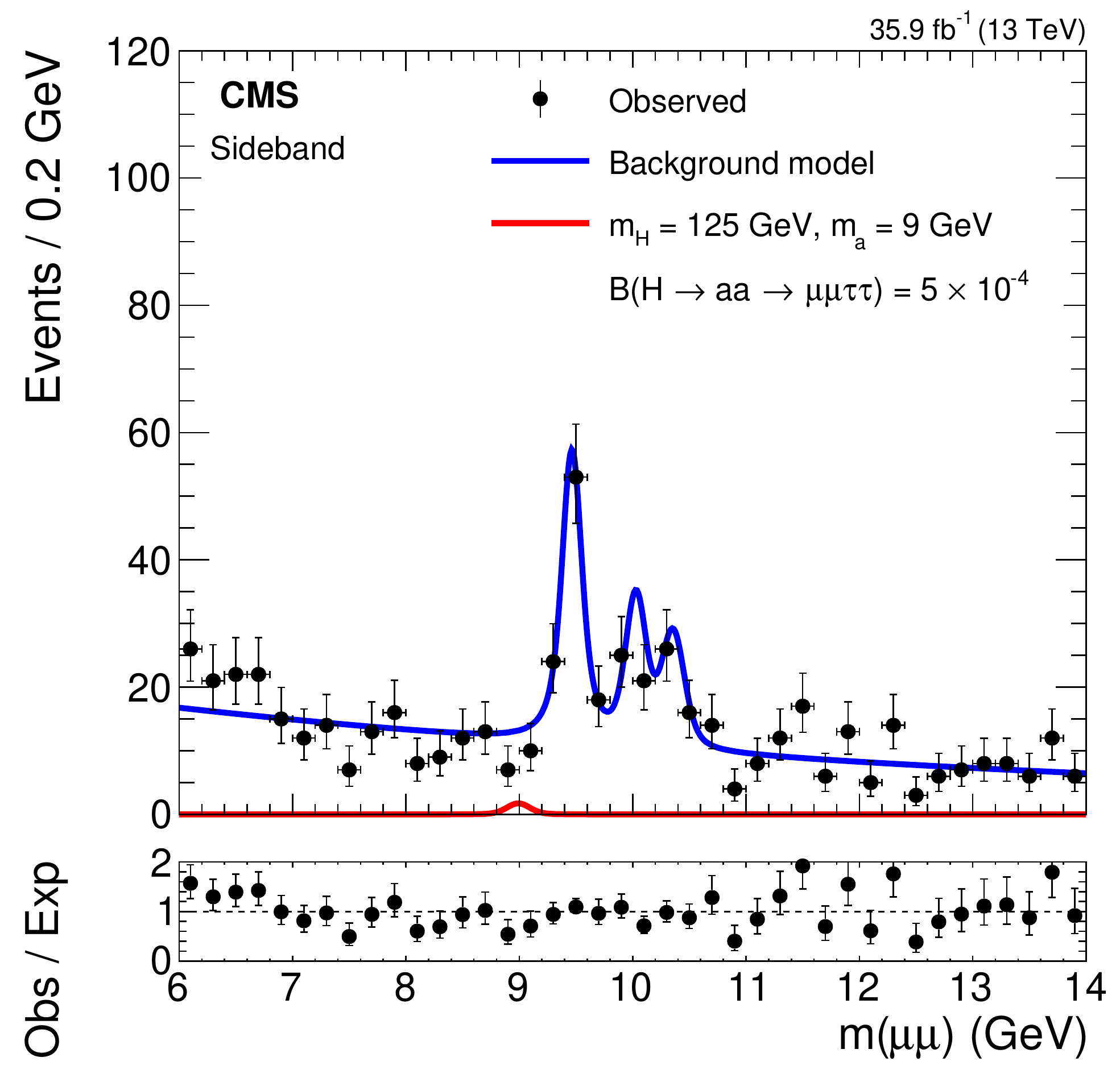}
  \includegraphics[width=0.4\textwidth]{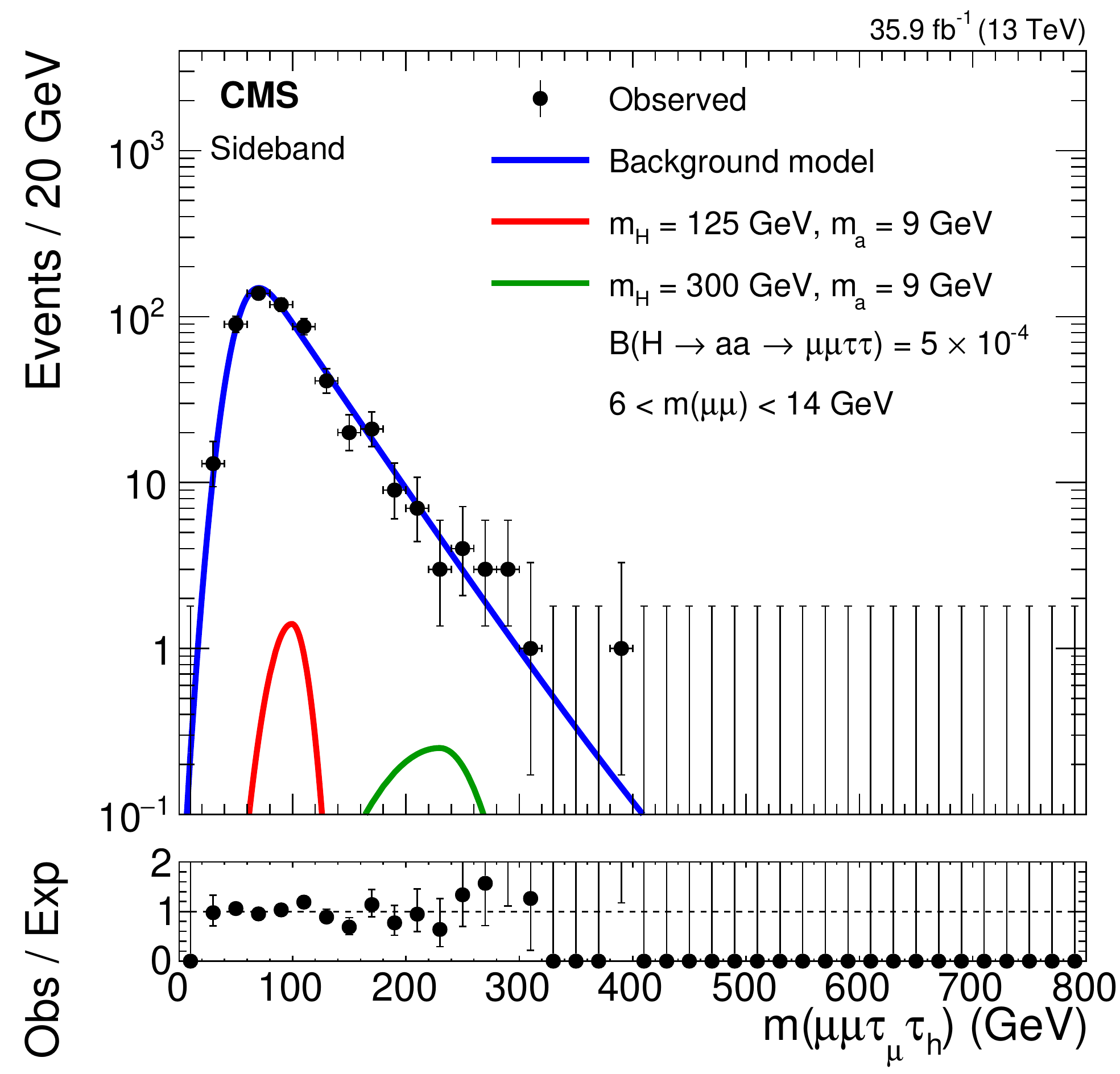} \\
  \includegraphics[width=0.4\textwidth]{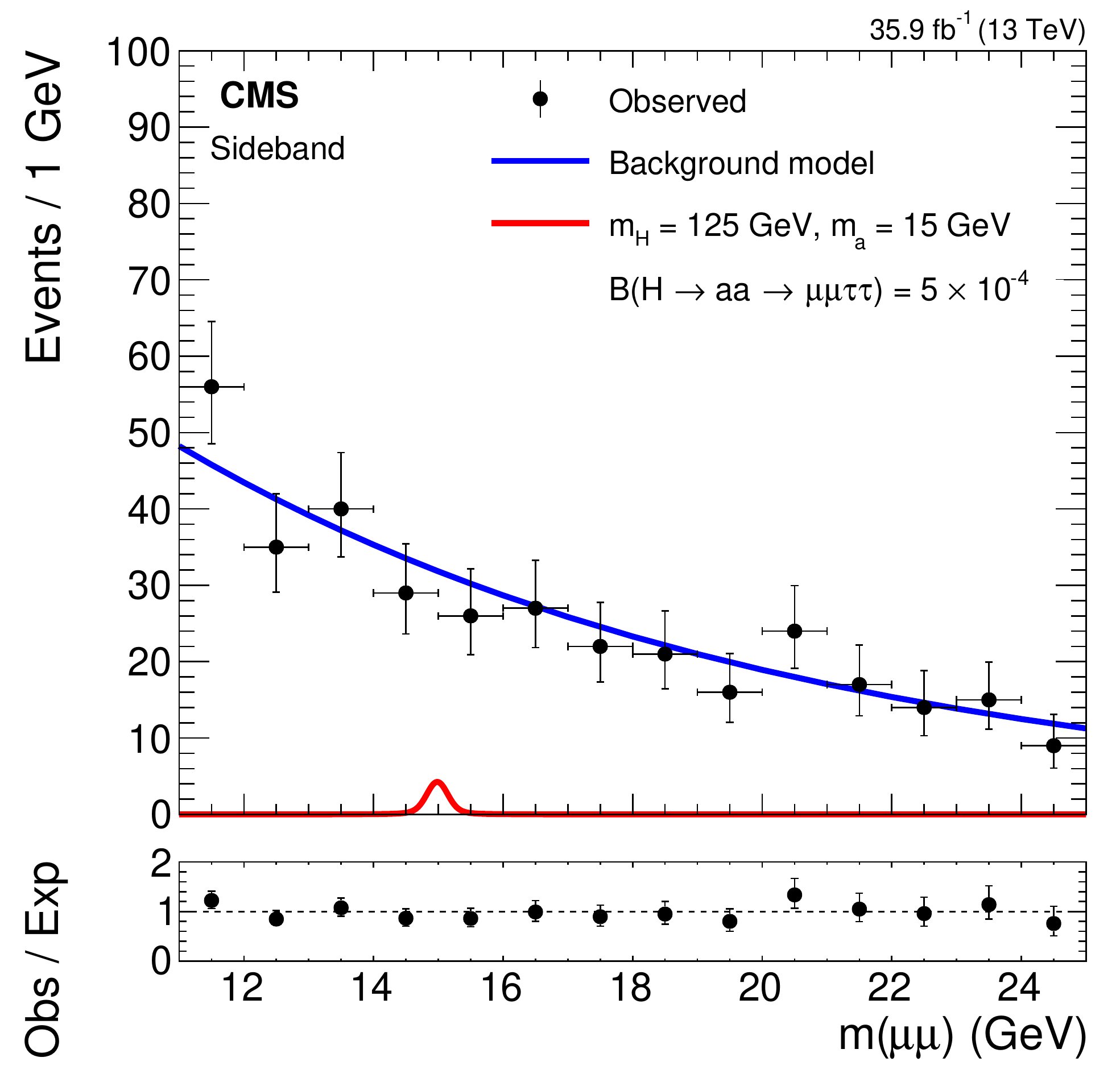}
  \includegraphics[width=0.4\textwidth]{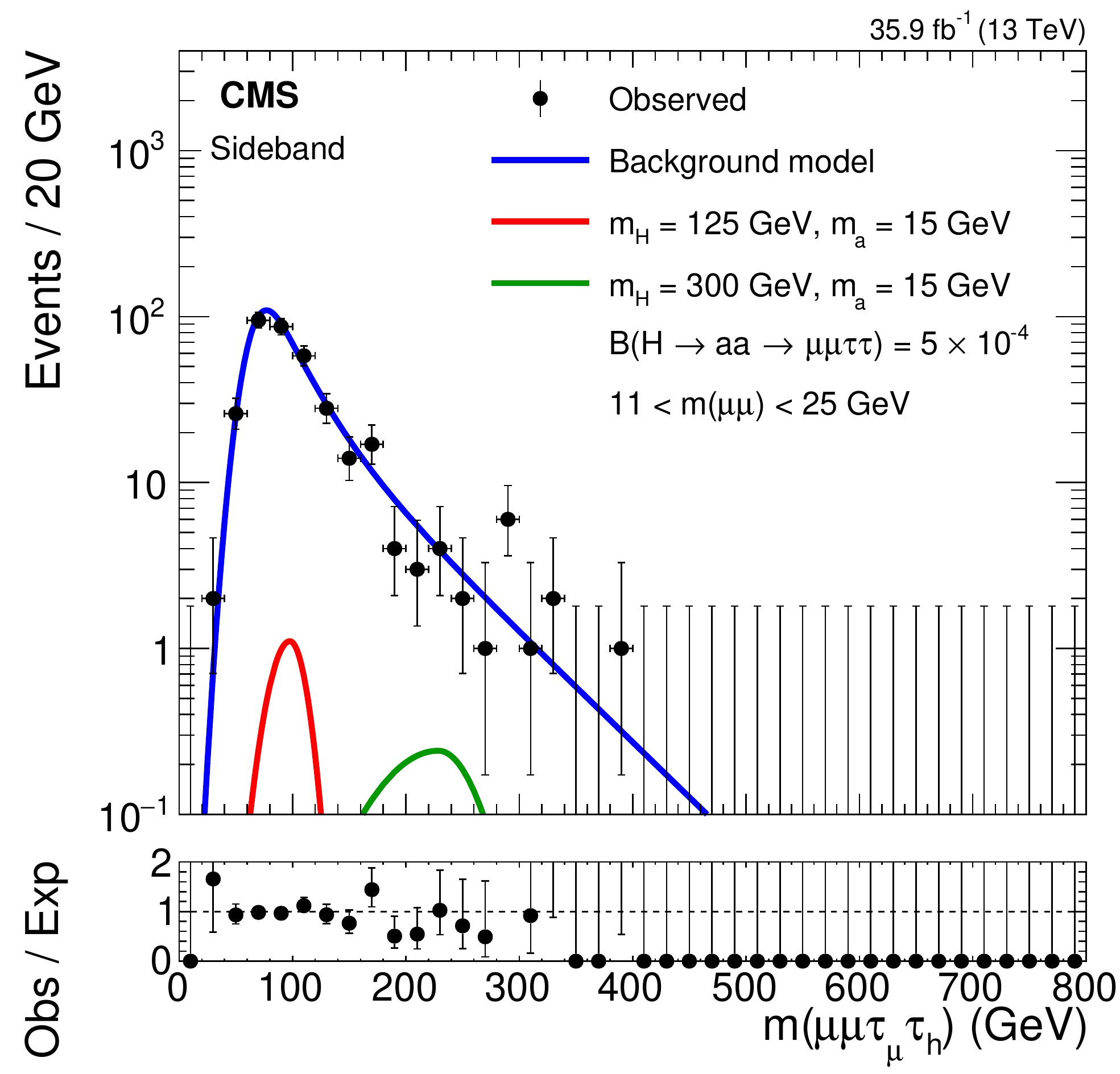}
  \caption{
  Projections of 2D background model fits and observed data in the sideband
  on the \mMM (left),
  and \mMMTT (right) axes
  with sample signal distributions that assume \PH boson masses of $\mPH=125$ and 300\GeV.
  The figures are divided into three fit ranges:
  $2.5<\mMM<8.5\GeV$ (upper), $6<\mMM<14\GeV$ (middle),
  and $11<\mMM<25\GeV$ (lower).
           }
  \label{fig:bgfitssideband}
\end{figure}

\begin{figure}[htbp]
  \centering
  \includegraphics[width=0.4\textwidth]{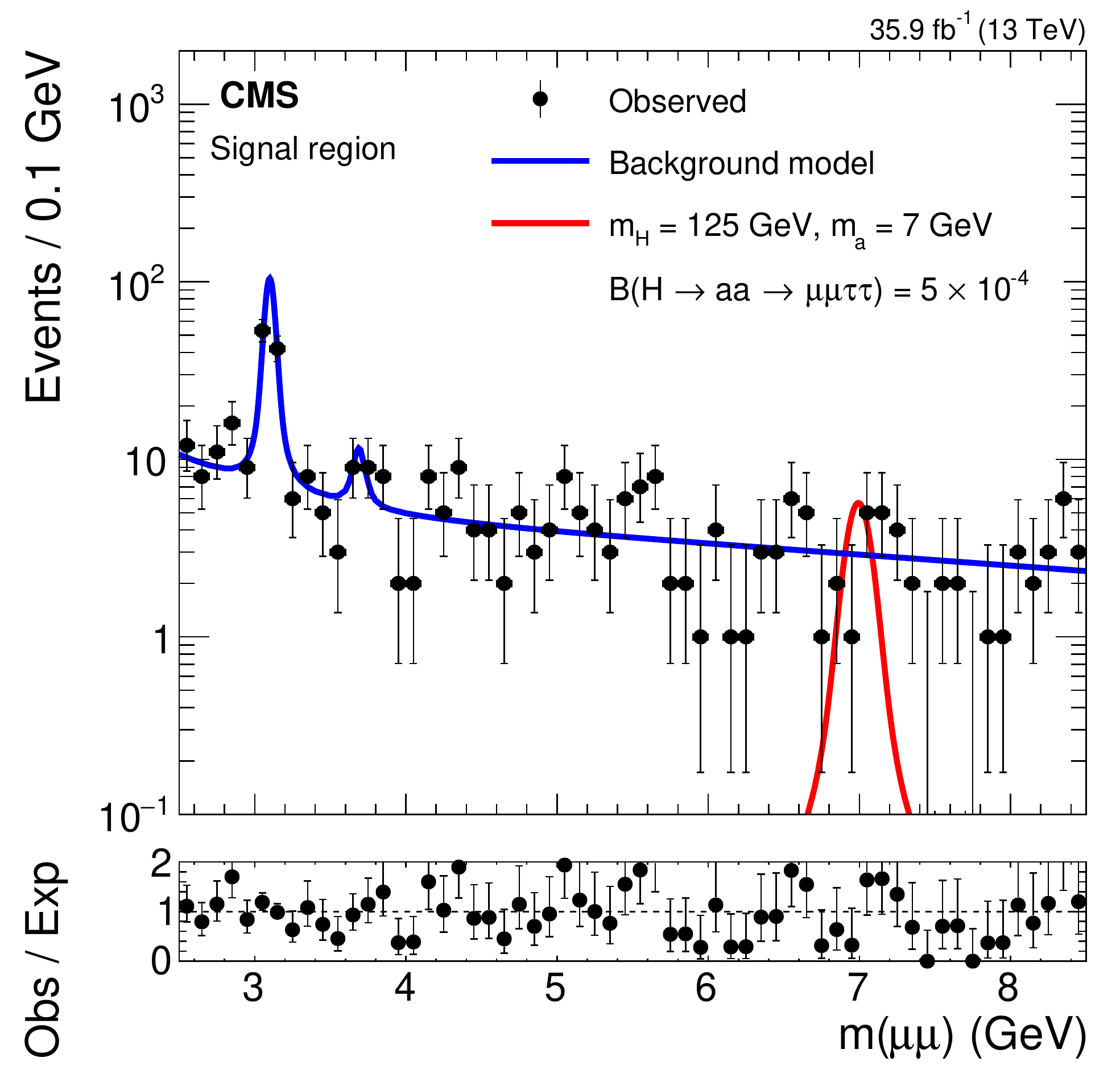}
  \includegraphics[width=0.4\textwidth]{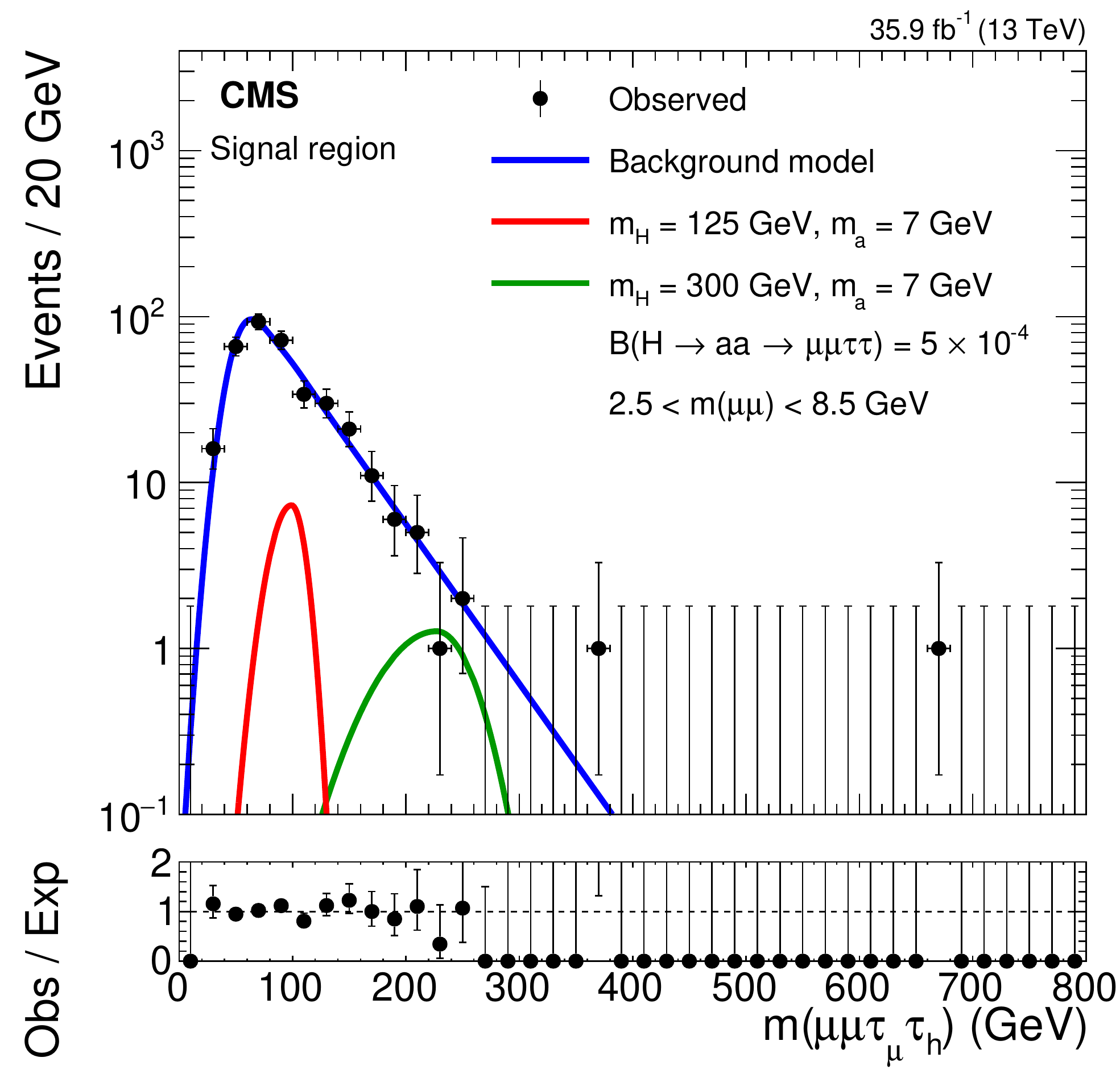} \\
  \includegraphics[width=0.4\textwidth]{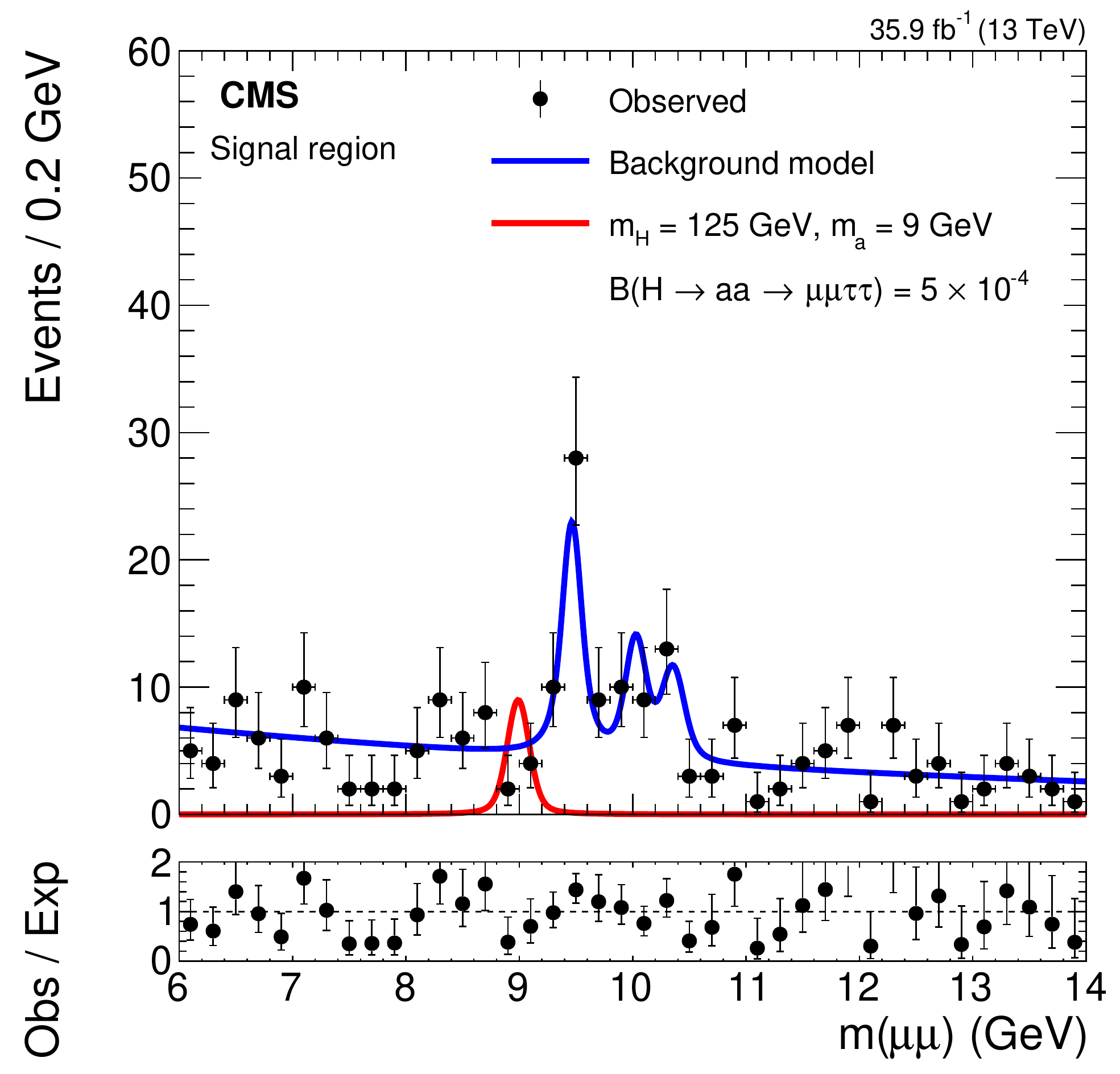}
  \includegraphics[width=0.4\textwidth]{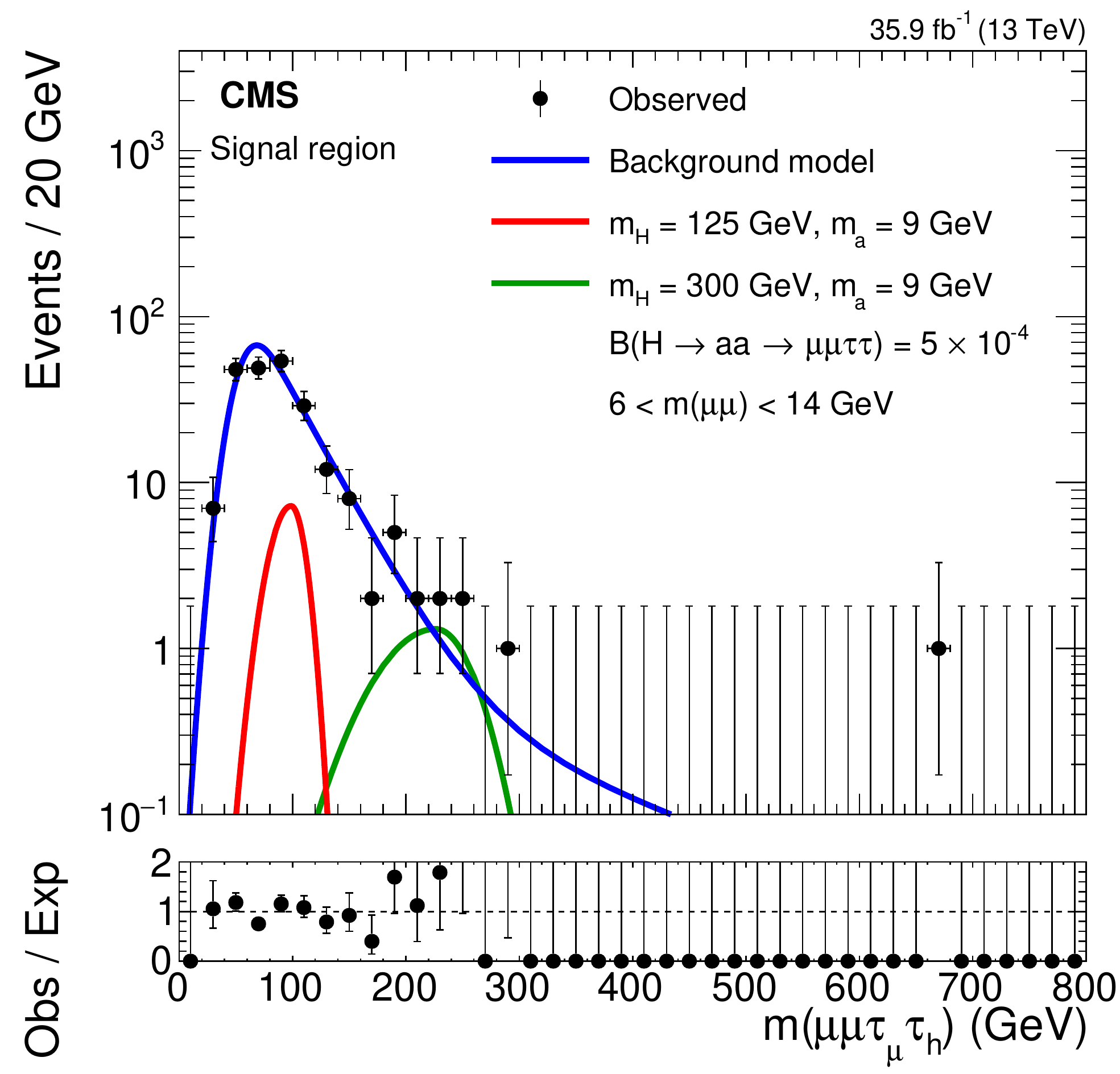} \\
  \includegraphics[width=0.4\textwidth]{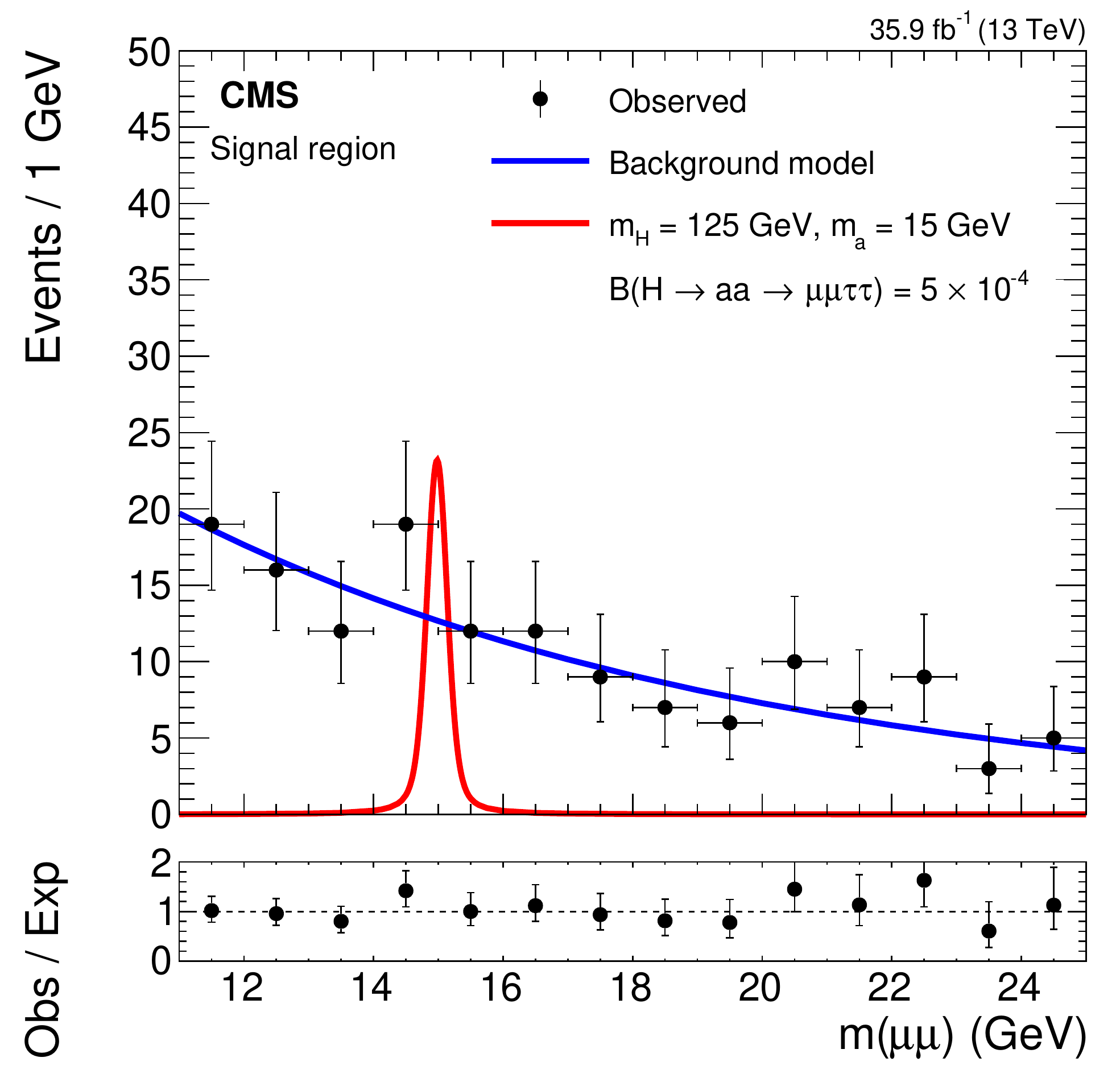}
  \includegraphics[width=0.4\textwidth]{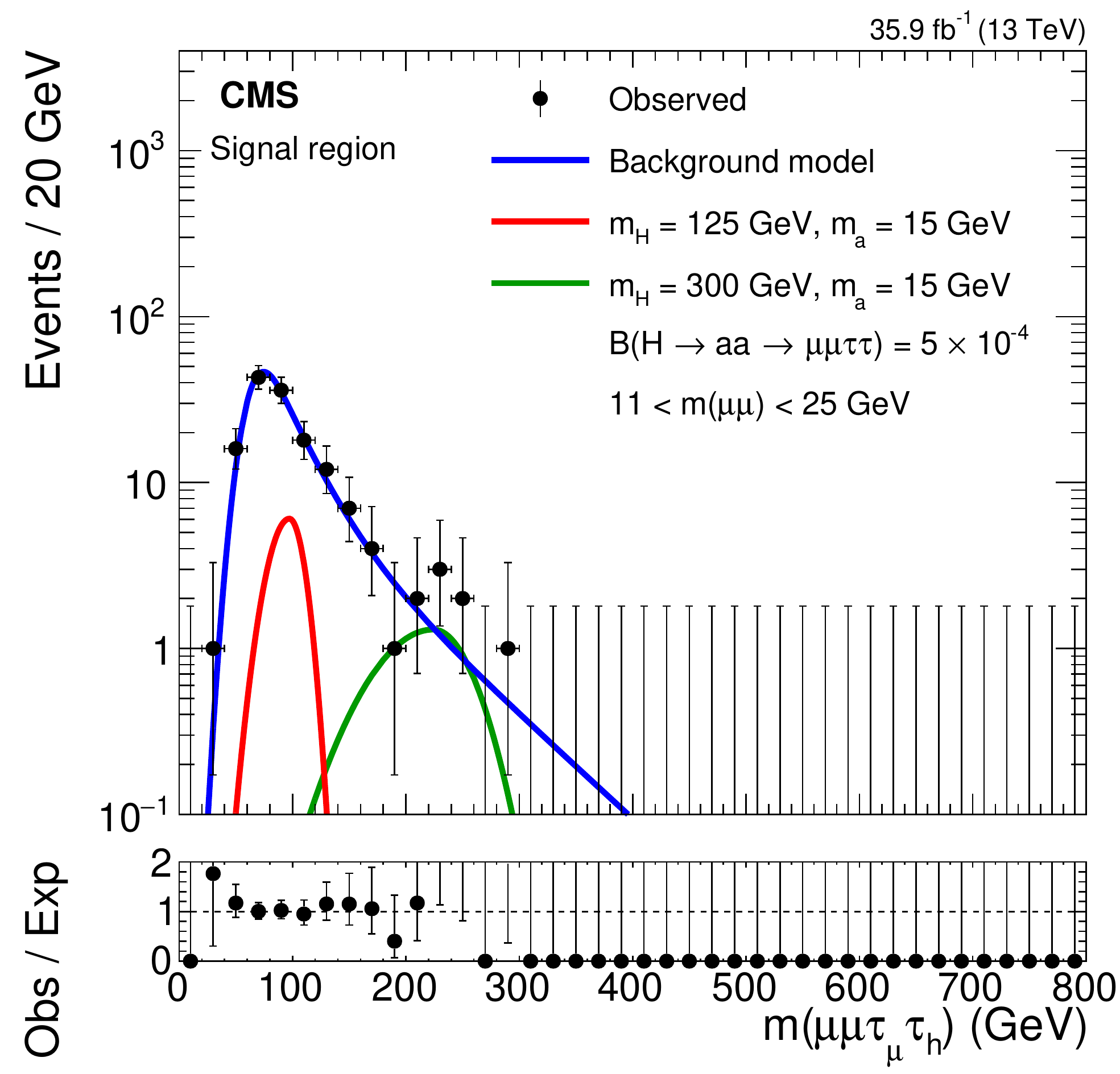}
  \caption{
  Projections of 2D background model fits and observed data in the signal region
  on the \mMM (left),
  and \mMMTT (right) axes
  with sample signal distributions that assume \PH boson masses of $\mPH=125$ and 300\GeV.
  The figures are divided into three fit ranges:
  $2.5<\mMM<8.5\GeV$ (upper), $6<\mMM<14\GeV$ (middle),
  and $11<\mMM<25\GeV$ (lower).
           }
  \label{fig:bgfits}
\end{figure}

The background model and observed data in the control region are shown
in Fig.~\ref{fig:bgfitscontrol}.
Projections on the \mMM and \mMMTT axes of the 2D background model
and observed data with sample signal distributions
for each fit range
are shown in Figs.~\ref{fig:bgfitssideband} and \ref{fig:bgfits} for the sideband and signal region, respectively.  
The signal distribution is scaled assuming an SM Higgs boson production cross section~\cite{deFlorian:2016spz} 
and $\mathB(\PH\to\Pa\Pa\to\PGm\PGm\PGt\PGt)=5\times10^{-4}$.
A small level of signal contamination is expected in the sideband and is included in the fit.
For the signal processes, there is minimal correlation between the
\mMM and the \mMMTT distributions.


\section{Systematic uncertainties}
\label{sec:unc}

Uncertainties in the signal process modeling contribute both
to the total expected signal yield and the individual signal fit
parameters.  Despite the small spatial separation between the \PGtmu and \PGth
candidates, the $\PGtmu\PGth$ reconstruction procedure, which relies on the excellent muon
discrimination of the CMS detector, allows the uncertainties in
the \PGth efficiency and energy scale modeling to be treated
independently from those for the \PGtmu candidates.
Systematic uncertainties in the efficiency measurements
from the tag-and-probe technique contribute
an uncertainty in the total signal yield of
0.5\% for the muon trigger efficiency and
1.0--1.4\% for each reconstructed muon.
The uncertainty in the muon momentum scale is 0.2--5.0\%;
most muons have $\pt<100\GeV$ and thus an uncertainty of 0.2\%~\cite{Sirunyan:2018fpa}.
For the \PGth reconstruction, there is an uncertainty in the \PGth identification efficiency of 5--18\%,
varying with $\pt(\PGth)$,
and an uncertainty in the \PGth energy scale of 1.2--3.0\%~\cite{Sirunyan:2018pgf}, 
varying with the number of charged and neutral hadrons in the \PGth decay.

The uncertainty
in the luminosity normalization of simulated signal samples is
2.5\%~\cite{CMS-PAS-LUM-17-001}.  
Uncertainty from pileup effects arises from
the uncertainty of 4.6\%~\cite{Sirunyan:2018nqx} in the total inelastic cross section of $\Pp\Pp$ interactions
resulting in a 1\% uncertainty in the signal yields.
The efficiency correction for the rejection of
jets tagged as originating from {\cPqb} quarks
contributes an uncertainty of up to 3\% in the signal yield.

As described in
Section~\ref{sec:samples}, a correction to the simulated ggF signal samples
to account for small differences in acceptance for the ggF and VBF
\PH boson production modes contributes a 0.5\% uncertainty in the signal
yield.
Theoretical uncertainties in the \PH boson production cross section
are calculated by varying renormalization ($\mu_\text{R}$)
and factorization ($\mu_\text{F}$) scales independently 
up and down by a factor of two with respect to the default values
with the condition that $0.5\leq\mu_\text{R}/\mu_\text{F}\leq2$.
The resulting uncertainties, combined with those
from Ref.~\cite{deFlorian:2016spz}, contribute less than 1\% to
the overall signal yield uncertainty.

For the background model, the tight-to-loose method
contributes a 15\% uncertainty in the total expected yield in the signal region.
This uncertainty arises from the application of the tight-to-loose
ratio to the validation sideband to obtain a prediction for the model shapes
in the validation region.
The additional uncertainty in the relative normalizations
of the low-mass meson resonances
arises from differences in the tight-to-loose method
predictions of the signal region distributions
when derived from the sideband,
as discussed in Section~\ref{sec:bkg}.
This uncertainty is measured to be 5--20\% for
\PGyP{2S} and each $\Upsilon$ resonance,
which yields up to a 3\% uncertainty near these
resonances in the final result.

\section{Results}
\label{sec:results}

\begin{figure}
  \centering
  \includegraphics[scale=0.4]{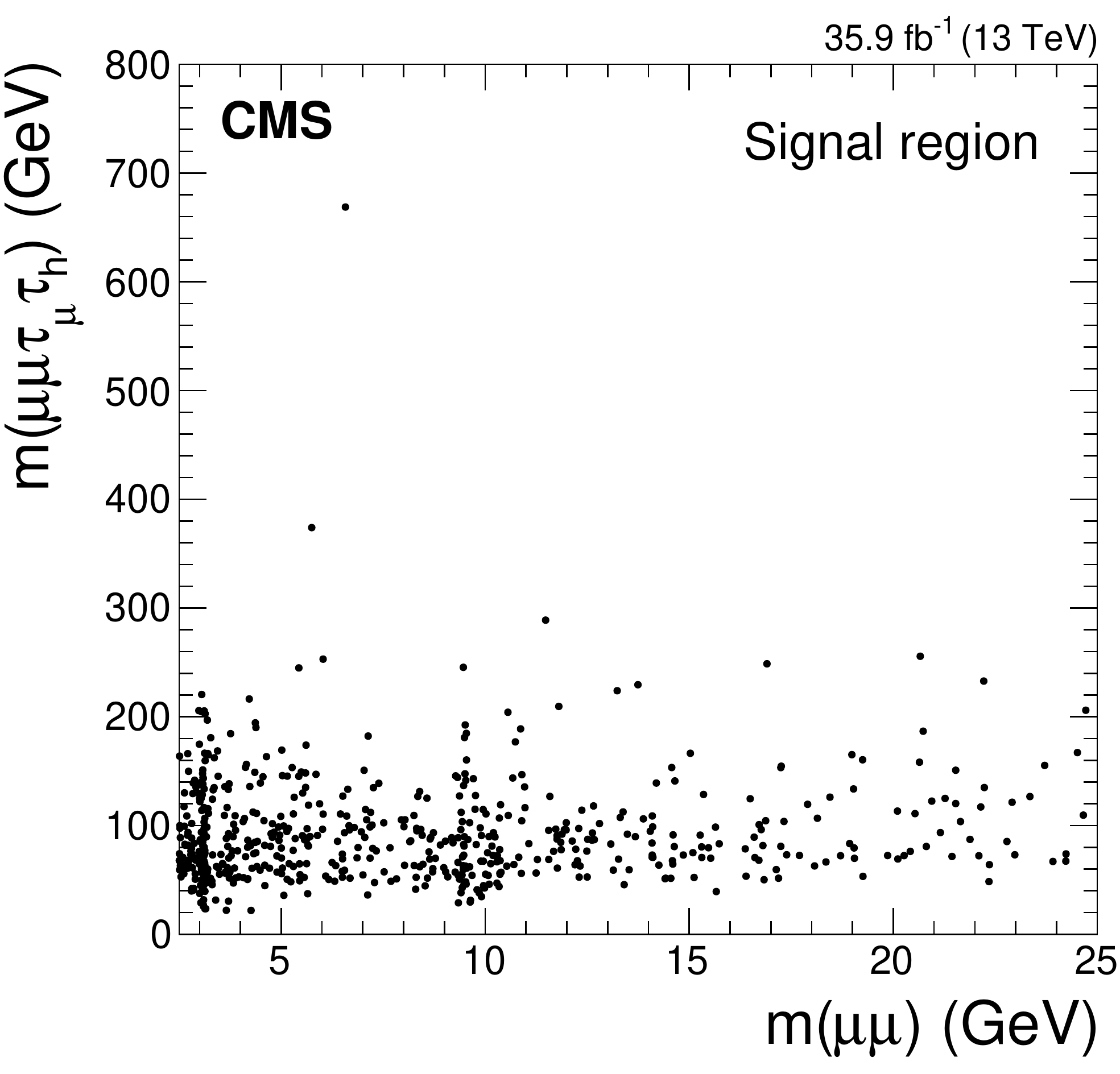}
  \caption{
Observed data distribution, as a function of the 4-body visible mass and $\PGm\PGm$ invariant mass
for the signal region; 614 events are observed.
          }
  \label{fig:dataObs}
\end{figure}

The observed distribution of data in the signal region is shown in
Figs.~\ref{fig:bgfits} and~\ref{fig:dataObs}.  No significant excess of events is observed above the expected SM
background.  
A modified frequentist approach based on the \CL criterion~\cite{Junk:1999kv,Read:2002hq} is used for upper limit calculations~\cite{PDG2018}
using the LHC test statistic~\cite{CMS-NOTE-2011-005}.
Systematic uncertainties are represented as nuisance parameters
assuming a log-normal PDF in the likelihood fit for uncertainties in the expected yields
and a Gaussian PDF of uncertainties in the signal and background model parameters.

\begin{figure}[htbp]
  \centering
  \includegraphics[width=0.4\textwidth]{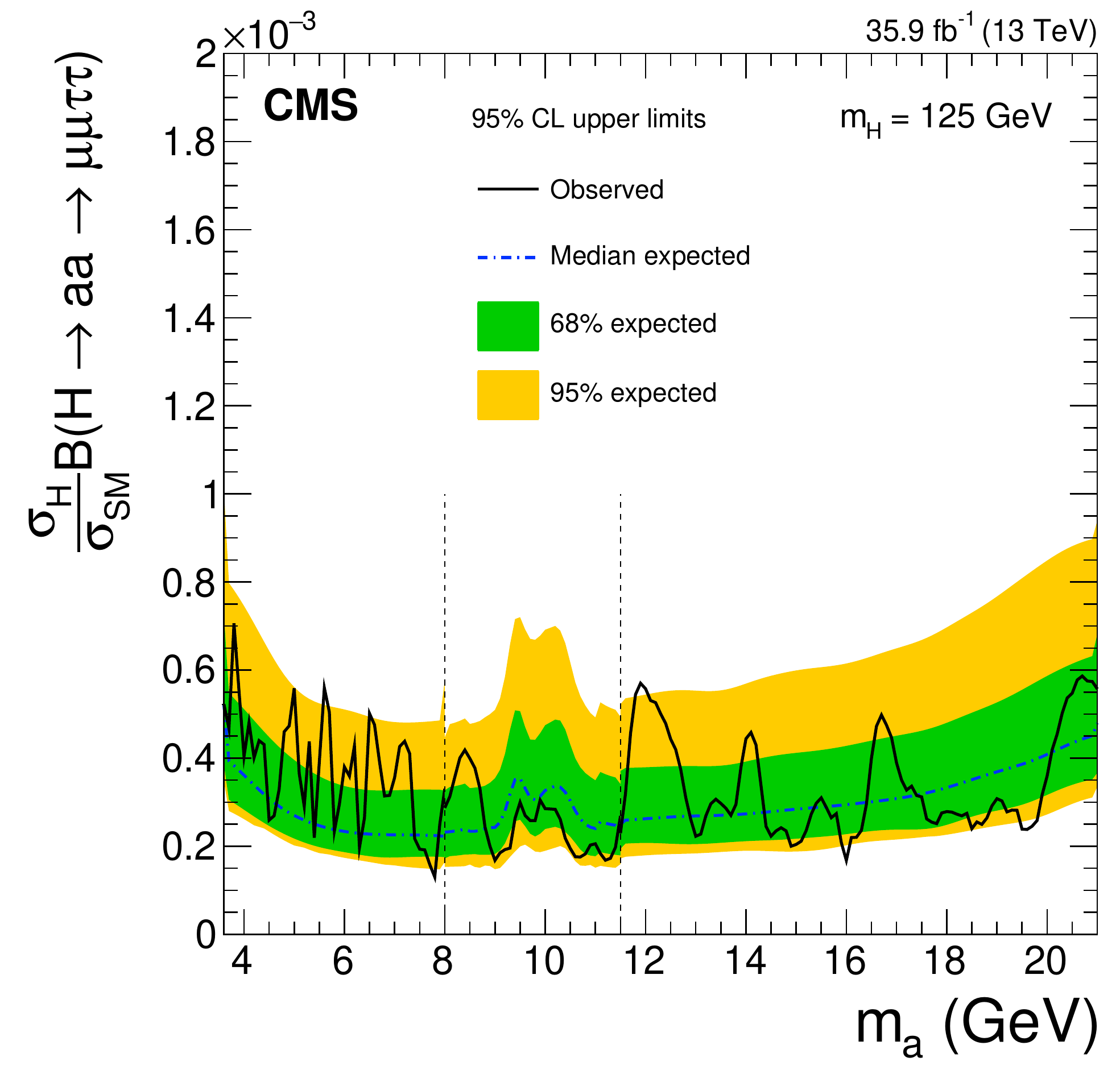}
  \includegraphics[width=0.4\textwidth]{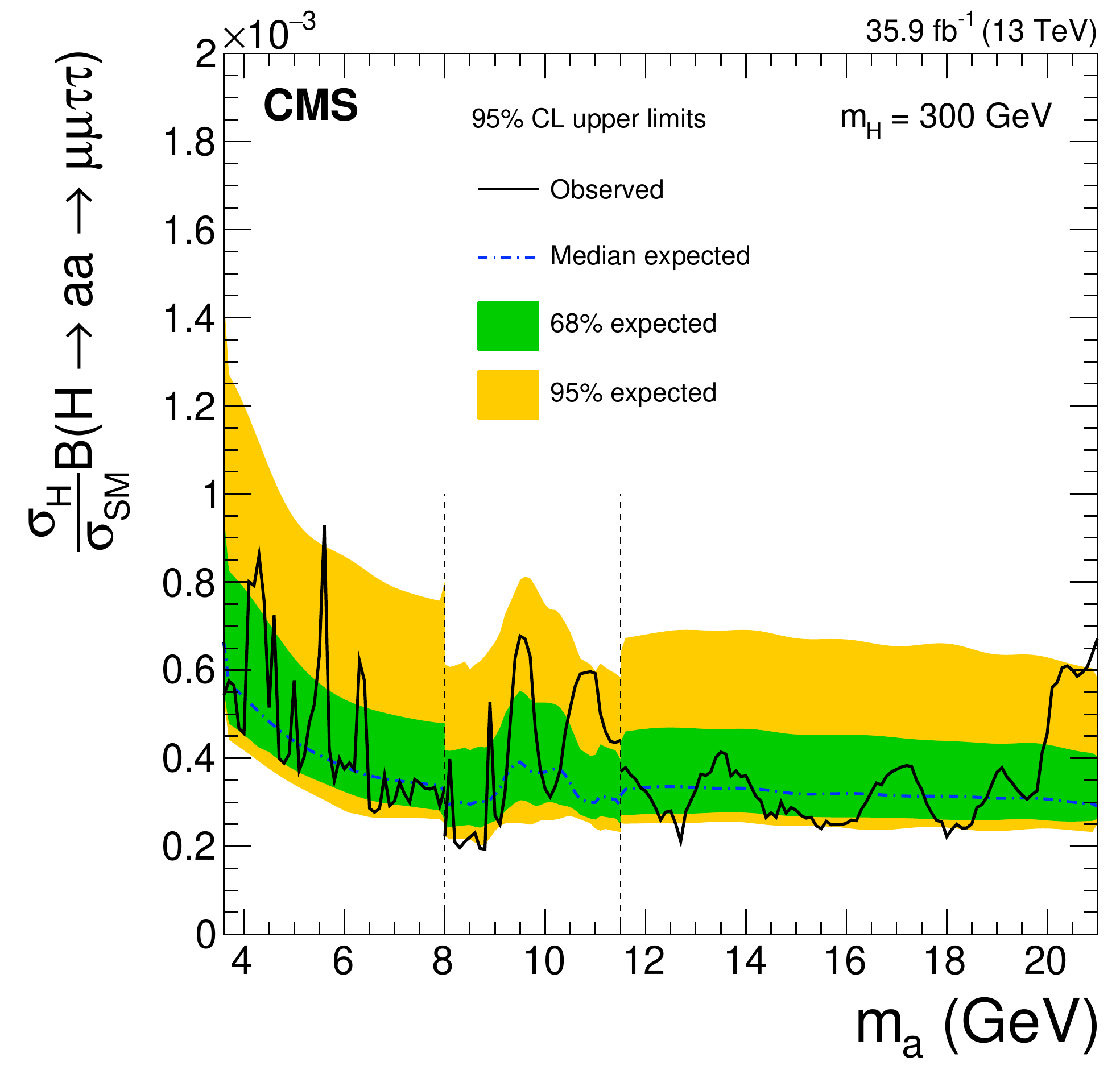}
  \caption{
           Model-independent 95\% \CL upper limits on $\sPH\mathB(\PH\to\Pa\Pa\to\PGm\PGm\PGt\PGt)/\sSM$
           as a function of pseudoscalar boson mass for a Higgs boson with $\mPH=125\GeV$ (left),
           and 300\GeV (right).
           The vertical dashed lines indicate the transition between the $\PGm\PGm$ mass fit ranges
           for a given mass hypothesis, occurring at $\mPGa=8$ and 11.5\GeV.
           The inner (green) band and the outer (yellow) band indicate the regions containing 68 and 95\%, respectively, 
           of the distribution of limits expected under the background-only hypothesis.
           }
  \label{fig:limits}
\end{figure}

Model-independent upper limits at 95\% \CL are set on $\sPH\mathB(\PH\to\Pa\Pa\to\PGm\PGm\PGt\PGt)/\sSM$
and are presented in Fig.~\ref{fig:limits}.    Here, \sSM is the SM Higgs boson (or, for $\mPH=300\GeV$, \sSM is the SM-like Higgs boson) production cross section
including ggF and VBF production modes~\cite{deFlorian:2016spz}.  
Broadly, the sensitivity of this exclusion decreases at low values of \mPGa because of reconstruction
inefficiencies as the decay products of the $\PGt\PGt$ pair overlap.
In addition, for $\mPH=125\GeV$, as \mPGa increases, the Lorentz boost
decreases causing the products to be well separated, failing the requirement of $\DRTT<0.8$.
The two peaking structures around $\mPGa=10\GeV$ are from the $\Upsilon$ resonances
where the \PGUP{1S} resonance is resolvable but the \PGUP{2S} and \PGUP{3S} merge because the rejection power of the boosted $\PGtmu\PGth$ selection sufficiently reduces the number of events in and around these peaks.
A third peaking structure is not as apparent but is also present at the \PGyP{2S} resonance.
Comparison with an earlier $\sqrt{s}=13\TeV$ result from the CMS Collaboration~\cite{Sirunyan:2018mbx} 
targeting resolved $\PGt\PGt$ decay products is possible for SM Higgs boson decays with $15<\mPGa<21\GeV$.
In this case, the two approaches have similar sensitivity.

\begin{figure}[htbp]
  \centering
  \includegraphics[width=0.4\textwidth]{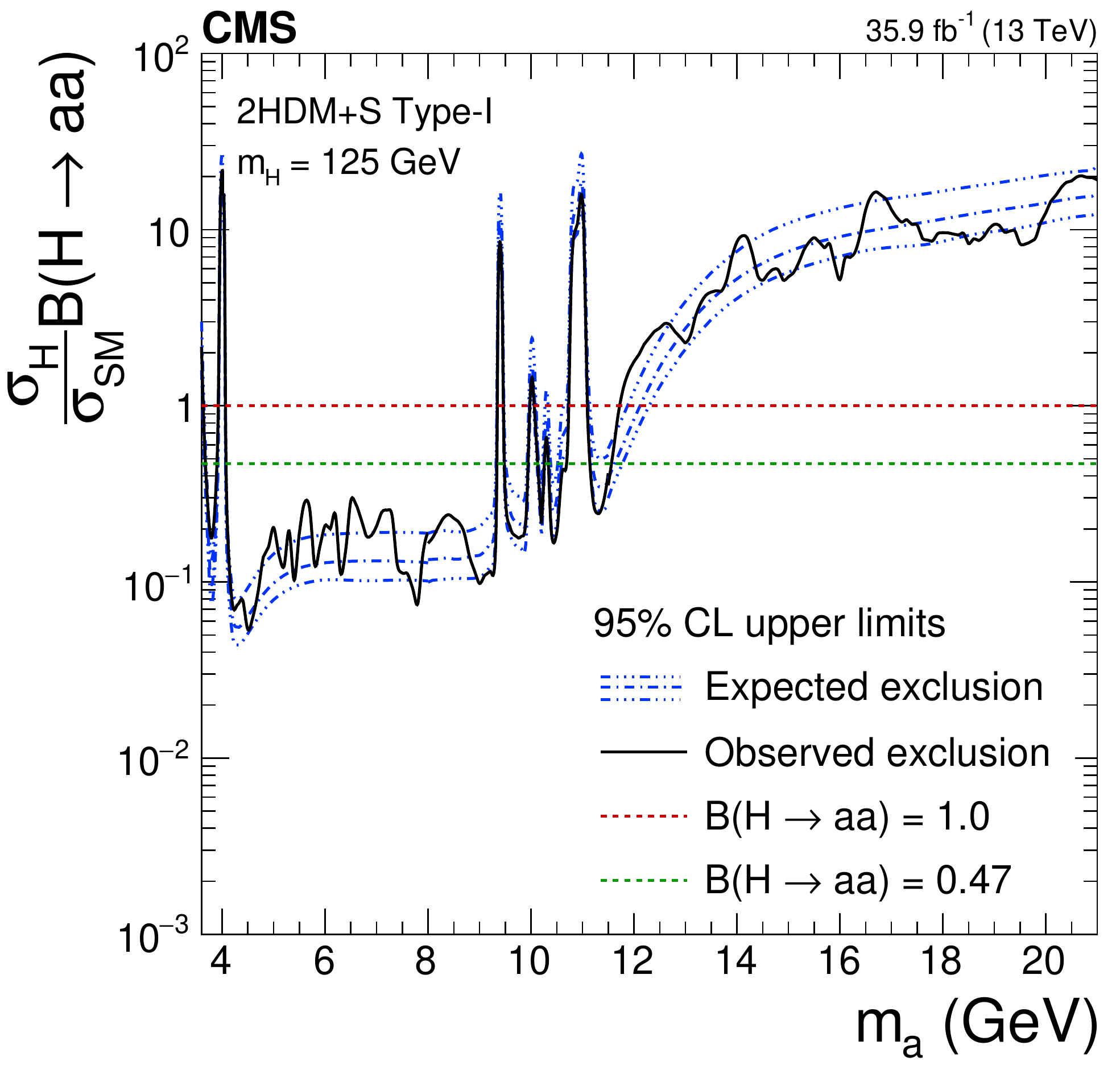}
  \caption{
      Observed (black) and expected (blue, median and 68\%) 
      model-specific 95\% \CL upper limits on $\sPH\mathB(\PH\to\Pa\Pa)/\sSM$
      as a function of \mPGa
      for the Type-I 2HDM+S at $\tanb = 1.5$
      and $\mPH=125\GeV$.  The assumed model branching fractions for pseudoscalar Higgs boson decay to $\PGm\PGm$
and $\PGt\PGt$ are taken from Ref.~\cite{Haisch:2018kqx} and are approximately independent of \tanb.
           }
  \label{fig:limits2HDMSProjection}
\end{figure}

\begin{figure}[htbp]
  \centering
  \includegraphics[width=0.4\textwidth]{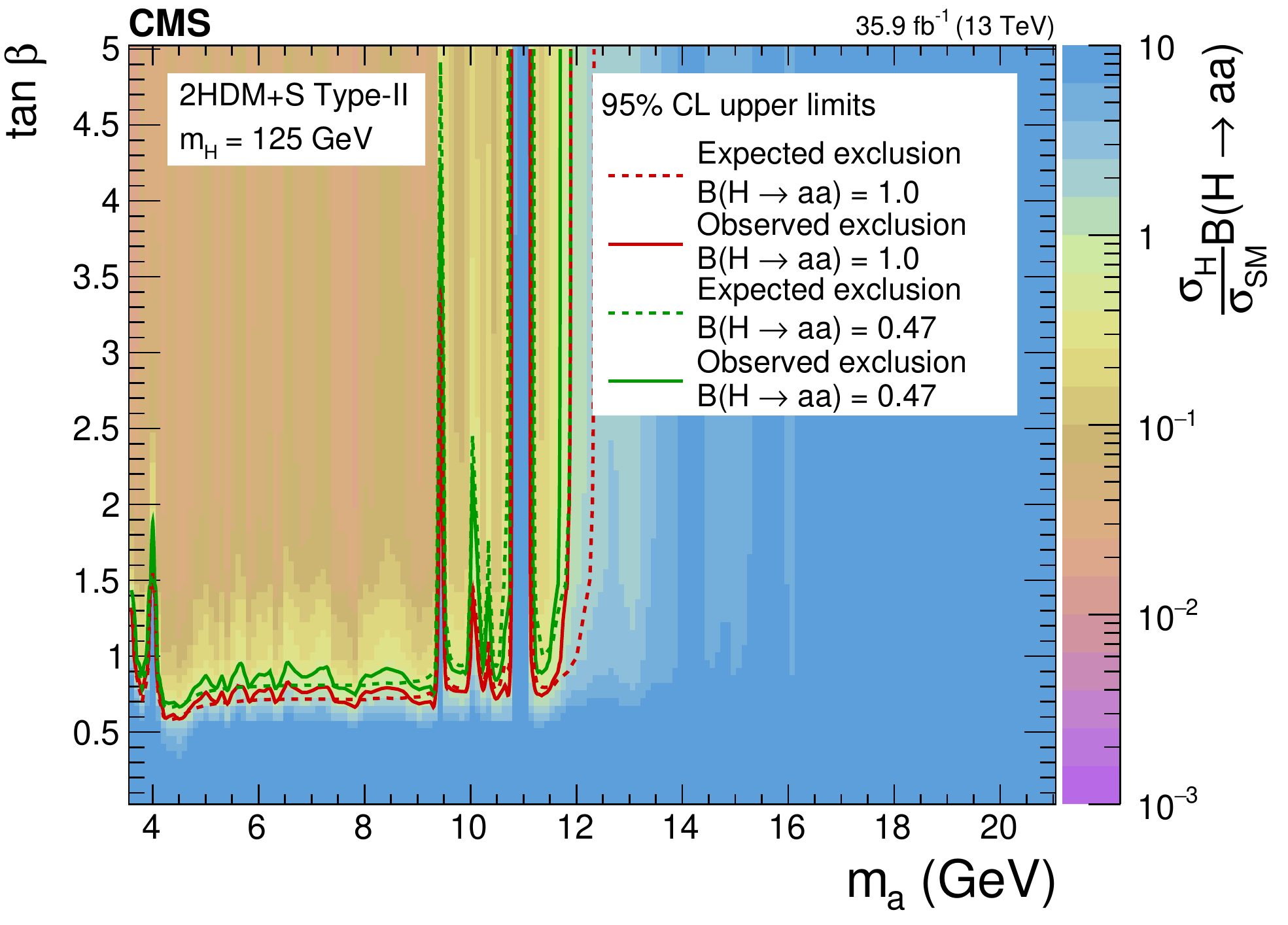} \\
  \includegraphics[width=0.4\textwidth]{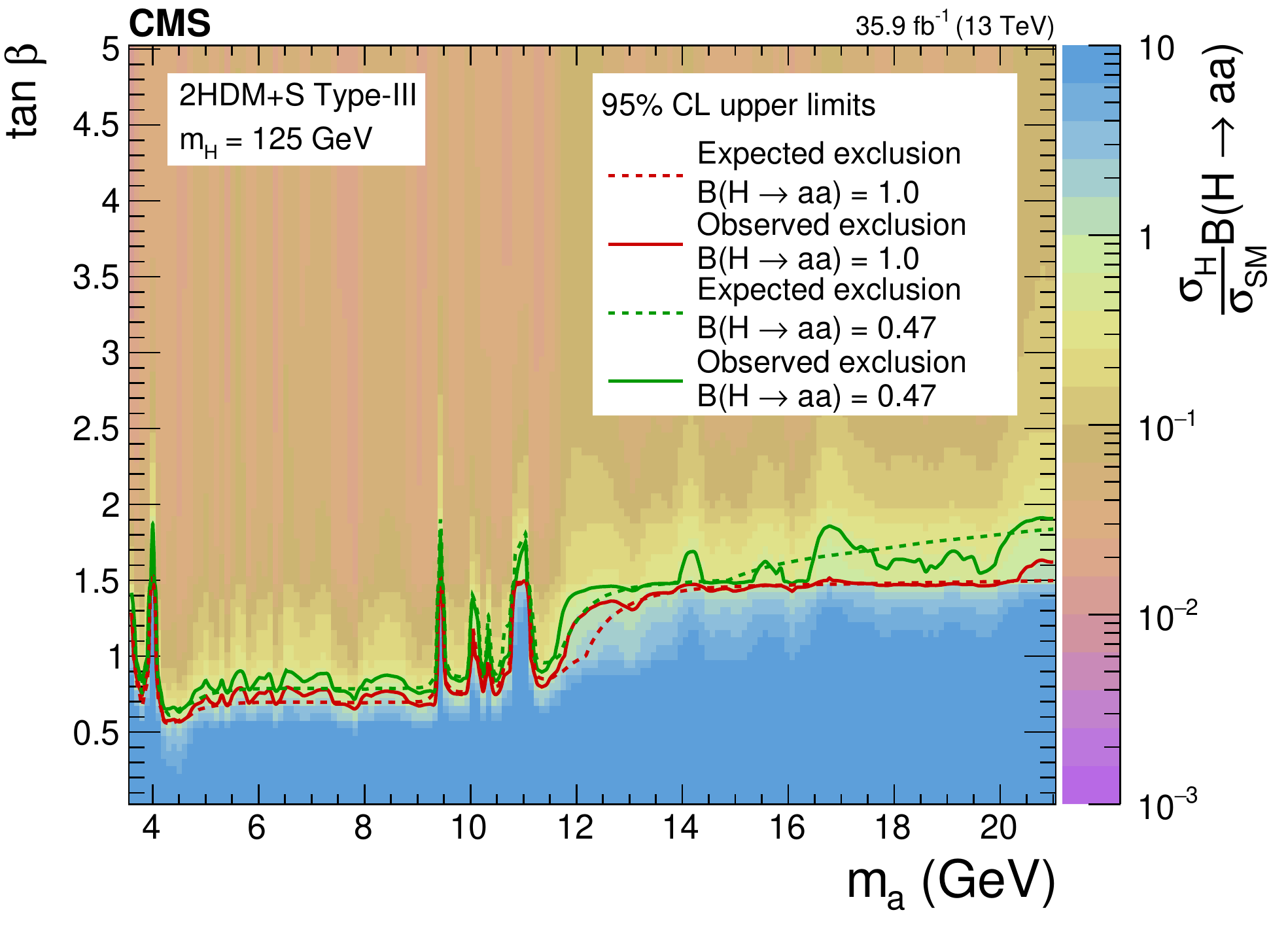}
  \includegraphics[width=0.4\textwidth]{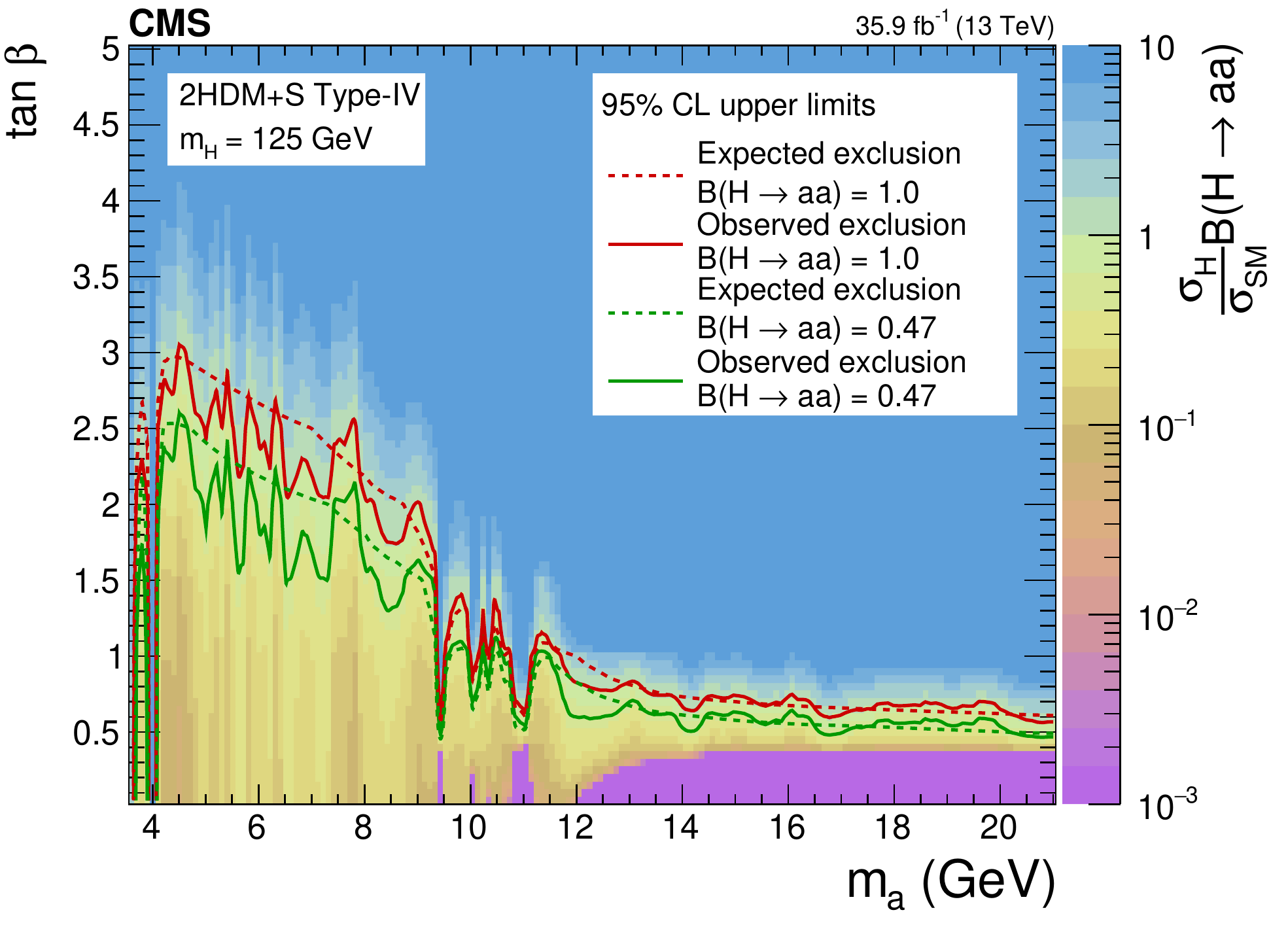}
  \caption{
      Model-specific 95\% \CL upper limits on $\sPH\mathB(\PH\to\Pa\Pa)/\sSM$
      for three model types of the 2HDM+S as a function of \tanb and \mPGa,
      for $\mPH=125\GeV$.  Contours for two values of $\mathB(\PH\to\Pa\Pa)$ are shown for reference.   The assumed model branching fractions for pseudoscalar Higgs boson decay to $\PGm\PGm$
and $\PGt\PGt$ are taken from Ref.~\cite{Haisch:2018kqx}.
           }
  \label{fig:limits2HDMS}
\end{figure}

Upper limits on $\sPH\mathB(\PH\to\Pa\Pa)/\sSM$ for the 2HDM+S for each Type-I to -IV as a function of
\tanb and \mPGa are shown in Figs.~\ref{fig:limits2HDMSProjection} and \ref{fig:limits2HDMS}.
The assumed model branching fractions for pseudoscalar decays to $\PGm\PGm$
and $\PGt\PGt$ are taken from Ref.~\cite{Haisch:2018kqx}, 
and the branching fraction $\mathB({\Pa}{\Pa}\to\PGm\PGm\PGt\PGt)$ depends strongly
on the 2HDM+S type~\cite{Curtin:2013fra}. 
The branching fractions are calculated in \tanb increments of 0.5 above $\tanb=1$
and increments of 0.1 below, and
a linear interpolation is applied between the calculated points in Fig.~\ref{fig:limits2HDMS}.
For the Type-I and -II models, we primarily probe the $2\mPGt<\mPGa<2\mPqb$ range,
with the Type-I upper limits approximately independent of \tanb.
In the Type-I model, the
most stringent limit of 5\% is set for $\mPGa\approx4.5\GeV$.
In the Type-III model, this analysis has exclusion power over the
full pseudoscalar mass range probed, especially at large \tanb.
For the
Type-II and -III models with \mPGa below the
$\bbbar$ threshold, upper limits on $\mathB(\PH\to\Pa\Pa)$
are stronger than the 0.47 inferred from combined measurements of SM
Higgs couplings~\cite{Sirunyan:2018koj} for $\tanb\ga0.8$-0.9, becoming as strong as 10\% for
$\tanb\ga1.5$.
In the Type-III models, strong upper limits are set for all pseudoscalar boson masses tested when $\tanb\ga1.5$.
The Type-IV model, however, can only be effectively probed 
in the low-\tanb region.  For a given \mPGa, the ratio of decay rates to $\PGm\PGm$ and $\PGt\PGt$, 
respectively $\mathB(\Pa\to\PGm\PGm)$ and $\mathB(\Pa\to\PGt\PGt)$,
depends only on \mPGm and \mPGt~\cite{Curtin:2013fra,Haisch:2018kqx}.  
Thus, these results can be converted into upper limits on $\sPH\mathB(\PH\to\Pa\Pa)/\sSM$.
Contours for different $\mathB(\PH\to\Pa\Pa)$ values are overlaid.
Compared with an earlier result by CMS~\cite{Sirunyan:2018mbx}, these upper limits are more stringent (where they can be compared) and extend to lower values of \mPGa.

\section{Summary}
\label{sec:summary}

A search for Higgs boson (\PH) decays to a pair of light pseudoscalar bosons (\Pa) is presented,
including the first such LHC results for an \PH with mass above 125\GeV.
The light pseudoscalars decay to $\PGm\PGm$ and $\PGt\PGt$
with substantial overlap between the leptons because of the Lorentz boost. 
This difficult topology motivates the development of a dedicated $\PGtmu\PGth$ reconstruction method
to increase the acceptance.
Data collected by the CMS Collaboration at $\sqrt{s}=13\TeV$, corresponding to an integrated luminosity of 35.9\fbinv, are examined
and no significant excess over standard model (SM) processes is observed.
This analysis obtains model-independent upper limits at 95\% confidence level on
the branching fraction (\mathB) of a SM-like Higgs boson (\PH),
decaying to a pair of pseudoscalar bosons (\Pa) in the $\PGm\PGm\PGt\PGt$ final state,
$\sPH\mathB(\PH\to\Pa\Pa\to\PGm\PGm\PGt\PGt)/\sSM$,
as well as model-specific upper limits on 
$\sPH\mathB(\PH\to\Pa\Pa)/\sSM$ 
for Type-I, -II, -III, and -IV two Higgs doublets plus singlet models.
In the Type-I model, the upper limit on the allowed branching fraction
is approximately independent of \tanb, with the
most stringent limit of 5\% set for $\mPGa\approx4.5\GeV$.
For the
Type-II and -III models with \mPGa below the
$\bbbar$ threshold, upper limits on $\mathB(\PH\to\Pa\Pa)$
are stronger than the 0.47 inferred from combined measurements of SM
Higgs couplings for $\tanb\ga0.8$-0.9, becoming as strong as 10\% for
$\tanb\ga1.5$.  In the Type-III models, the predicted branching fraction
to leptons increases with \tanb, leading to strong
upper limits for all pseudoscalar boson masses tested when $\tanb\ga1.5$.  In
contrast, the strongest upper limits for Type-IV models are set when $\tanb<1$.
These results significantly extend upper limits obtained by earlier searches by the CMS and ATLAS Collaborations,
such as those obtained by CMS with 8\TeV data~\cite{Khachatryan:2017mnf}, 
and are complementary to present searches (\eg Ref.~\cite{Sirunyan:2018mbx}) at higher \mPGa that
lead to resolved $\PGm\PGm$ and $\PGt\PGt$ final states.

\begin{acknowledgments}
  We congratulate our colleagues in the CERN accelerator departments for the excellent performance of the LHC and thank the technical and administrative staffs at CERN and at other CMS institutes for their contributions to the success of the CMS effort. In addition, we gratefully acknowledge the computing centers and personnel of the Worldwide LHC Computing Grid for delivering so effectively the computing infrastructure essential to our analyses. Finally, we acknowledge the enduring support for the construction and operation of the LHC and the CMS detector provided by the following funding agencies: BMBWF and FWF (Austria); FNRS and FWO (Belgium); CNPq, CAPES, FAPERJ, FAPERGS, and FAPESP (Brazil); MES (Bulgaria); CERN; CAS, MoST, and NSFC (China); COLCIENCIAS (Colombia); MSES and CSF (Croatia); RPF (Cyprus); SENESCYT (Ecuador); MoER, ERC IUT, PUT and ERDF (Estonia); Academy of Finland, MEC, and HIP (Finland); CEA and CNRS/IN2P3 (France); BMBF, DFG, and HGF (Germany); GSRT (Greece); NKFIA (Hungary); DAE and DST (India); IPM (Iran); SFI (Ireland); INFN (Italy); MSIP and NRF (Republic of Korea); MES (Latvia); LAS (Lithuania); MOE and UM (Malaysia); BUAP, CINVESTAV, CONACYT, LNS, SEP, and UASLP-FAI (Mexico); MOS (Montenegro); MBIE (New Zealand); PAEC (Pakistan); MSHE and NSC (Poland); FCT (Portugal); JINR (Dubna); MON, RosAtom, RAS, RFBR, and NRC KI (Russia); MESTD (Serbia); SEIDI, CPAN, PCTI, and FEDER (Spain); MOSTR (Sri Lanka); Swiss Funding Agencies (Switzerland); MST (Taipei); ThEPCenter, IPST, STAR, and NSTDA (Thailand); TUBITAK and TAEK (Turkey); NASU (Ukraine); STFC (United Kingdom); DOE and NSF (USA).

  \hyphenation{Rachada-pisek} Individuals have received support from the Marie-Curie program and the European Research Council and Horizon 2020 Grant, contract Nos.\ 675440, 752730, and 765710 (European Union); the Leventis Foundation; the A.P.\ Sloan Foundation; the Alexander von Humboldt Foundation; the Belgian Federal Science Policy Office; the Fonds pour la Formation \`a la Recherche dans l'Industrie et dans l'Agriculture (FRIA-Belgium); the Agentschap voor Innovatie door Wetenschap en Technologie (IWT-Belgium); the F.R.S.-FNRS and FWO (Belgium) under the ``Excellence of Science -- EOS" -- be.h project n.\ 30820817; the Beijing Municipal Science \& Technology Commission, No. Z191100007219010; the Ministry of Education, Youth and Sports (MEYS) of the Czech Republic; the Deutsche Forschungsgemeinschaft (DFG) under Germany's Excellence Strategy -- EXC 2121 ``Quantum Universe" -- 390833306; the Lend\"ulet (``Momentum") Program and the J\'anos Bolyai Research Scholarship of the Hungarian Academy of Sciences, the New National Excellence Program \'UNKP, the NKFIA research grants 123842, 123959, 124845, 124850, 125105, 128713, 128786, and 129058 (Hungary); the Council of Science and Industrial Research, India; the HOMING PLUS program of the Foundation for Polish Science, cofinanced from European Union, Regional Development Fund, the Mobility Plus program of the Ministry of Science and Higher Education, the National Science Center (Poland), contracts Harmonia 2014/14/M/ST2/00428, Opus 2014/13/B/ST2/02543, 2014/15/B/ST2/03998, and 2015/19/B/ST2/02861, Sonata-bis 2012/07/E/ST2/01406; the National Priorities Research Program by Qatar National Research Fund; the Ministry of Science and Education, grant no. 14.W03.31.0026 (Russia); the Tomsk Polytechnic University Competitiveness Enhancement Program and ``Nauka" Project FSWW-2020-0008 (Russia); the Programa Estatal de Fomento de la Investigaci{\'o}n Cient{\'i}fica y T{\'e}cnica de Excelencia Mar\'{\i}a de Maeztu, grant MDM-2015-0509 and the Programa Severo Ochoa del Principado de Asturias; the Thalis and Aristeia programs cofinanced by EU-ESF and the Greek NSRF; the Rachadapisek Sompot Fund for Postdoctoral Fellowship, Chulalongkorn University and the Chulalongkorn Academic into Its 2nd Century Project Advancement Project (Thailand); the Kavli Foundation; the Nvidia Corporation; the SuperMicro Corporation; the Welch Foundation, contract C-1845; and the Weston Havens Foundation (USA). \end{acknowledgments}

\bibliography{auto_generated}
\cleardoublepage \appendix\section{The CMS Collaboration \label{app:collab}}\begin{sloppypar}\hyphenpenalty=5000\widowpenalty=500\clubpenalty=5000\input{HIG-18-024-authorlist.tex}\end{sloppypar}
\end{document}

%% file: HIG-18-024-authorlist.tex
\vskip\cmsinstskip
\textbf{Yerevan Physics Institute, Yerevan, Armenia}\\*[0pt]
A.M.~Sirunyan$^{\textrm{\dag}}$, A.~Tumasyan
\vskip\cmsinstskip
\textbf{Institut f\"{u}r Hochenergiephysik, Wien, Austria}\\*[0pt]
W.~Adam, F.~Ambrogi, T.~Bergauer, M.~Dragicevic, J.~Er\"{o}, A.~Escalante~Del~Valle, R.~Fr\"{u}hwirth\cmsAuthorMark{1}, M.~Jeitler\cmsAuthorMark{1}, N.~Krammer, L.~Lechner, D.~Liko, T.~Madlener, I.~Mikulec, N.~Rad, J.~Schieck\cmsAuthorMark{1}, R.~Sch\"{o}fbeck, M.~Spanring, S.~Templ, W.~Waltenberger, C.-E.~Wulz\cmsAuthorMark{1}, M.~Zarucki
% ...
\vskip\cmsinstskip
\dag: Deceased\\
1:  Also at Vienna University of Technology, Vienna, Austria\\
% ...


%% file: HIG-18-024_temp.bbl
\providecommand{\href}[2]{#2}\begingroup\raggedright\begin{thebibliography}{10}%
\makeatletter
\providecommand{\hrefCMSnoop }[0]{\@secondoftwo}%
\makeatother
\providecommand{\doi}{\texttt{doi:}\begingroup \urlstyle{tt}\Url}

\bibitem{Aad:2013wqa}
\hrefCMSnoop {}{{ATLAS Collaboration}, ``Measurements of {Higgs} boson
  production and couplings in diboson final states with the {ATLAS} detector at
  the {LHC}'',} \textit{ Phys. Lett. B} \textbf{ 726} (2013) 88,
  \href{http://dx.doi.org/10.1016/j.physletb.2013.08.010}{\doi{10.1016/j.physletb.2013.08.010}},
  \href{http://www.arXiv.org/abs/1307.1427}{\texttt{arXiv:1307.1427}}.
[Erratum: \DOI{10.1016/j.physletb.2014.05.011}].

\bibitem{Khachatryan:2016vau}
\hrefCMSnoop {}{{ATLAS and CMS Collaborations}, ``Measurements of the {Higgs}
  boson production and decay rates and constraints on its couplings from a
  combined {ATLAS} and {CMS} analysis of the {LHC} pp collision data at $
  \sqrt{s}=7 $ and 8 {TeV}'',} \textit{ JHEP} \textbf{ 08} (2016) 045,
  \href{http://dx.doi.org/10.1007/JHEP08(2016)045}{\doi{10.1007/JHEP08(2016)045}},
\href{http://www.arXiv.org/abs/1606.02266}{\texttt{arXiv:1606.02266}}.

\bibitem{Sirunyan:2017exp}
\hrefCMSnoop {}{{CMS Collaboration}, ``Measurements of properties of the
  {Higgs} boson decaying into the four-lepton final state in pp collisions at $
  \sqrt{s}=13 $ {TeV}'',} \textit{ JHEP} \textbf{ 11} (2017) 047,
  \href{http://dx.doi.org/10.1007/JHEP11(2017)047}{\doi{10.1007/JHEP11(2017)047}},
\href{http://www.arXiv.org/abs/1706.09936}{\texttt{arXiv:1706.09936}}.

\bibitem{Aad:2014aba}
\hrefCMSnoop {}{{ATLAS Collaboration}, ``Measurement of the {Higgs} boson mass
  from the {$H\rightarrow \gamma\gamma$ and $H \rightarrow ZZ^{*} \rightarrow
  4\ell$} channels with the {ATLAS} detector using 25 fb$^{-1}$ of pp collision
  data'',} \textit{ Phys. Rev. D} \textbf{ 90} (2014) 052004,
  \href{http://dx.doi.org/10.1103/PhysRevD.90.052004}{\doi{10.1103/PhysRevD.90.052004}},
\href{http://www.arXiv.org/abs/1406.3827}{\texttt{arXiv:1406.3827}}.

\bibitem{Aad:2015zhl}
\hrefCMSnoop {}{{ATLAS and CMS Collaborations}, ``Combined measurement of the
  {Higgs} boson mass in pp collisions at $\sqrt{s}=7$ and 8 {TeV} with the
  {ATLAS} and {CMS} experiments'',} \textit{ Phys. Rev. Lett.} \textbf{ 114}
  (2015) 191803,
  \href{http://dx.doi.org/10.1103/PhysRevLett.114.191803}{\doi{10.1103/PhysRevLett.114.191803}},
\href{http://www.arXiv.org/abs/1503.07589}{\texttt{arXiv:1503.07589}}.

\bibitem{Branco:2011iw}
G.~C. Branco\hrefCMSnoop {}{ {et~al.}, ``Theory and phenomenology of
  two-{Higgs}-doublet models'',} \textit{ Phys. Rept.} \textbf{ 516} (2012) 1,
  \href{http://dx.doi.org/10.1016/j.physrep.2012.02.002}{\doi{10.1016/j.physrep.2012.02.002}},
\href{http://www.arXiv.org/abs/1106.0034}{\texttt{arXiv:1106.0034}}.

\bibitem{Curtin:2013fra}
D.~Curtin\hrefCMSnoop {}{ {et~al.}, ``Exotic decays of the 125 {GeV} {Higgs}
  boson'',} \textit{ Phys. Rev. D} \textbf{ 90} (2014) 075004,
  \href{http://dx.doi.org/10.1103/PhysRevD.90.075004}{\doi{10.1103/PhysRevD.90.075004}},
\href{http://www.arXiv.org/abs/1312.4992}{\texttt{arXiv:1312.4992}}.

\bibitem{Ellwanger:2009dp}
\hrefCMSnoop {}{U.~Ellwanger, C.~Hugonie, and A.~M. Teixeira, ``The
  next-to-minimal supersymmetric standard model'',} \textit{ Phys. Rept.}
  \textbf{ 496} (2010) 1,
  \href{http://dx.doi.org/10.1016/j.physrep.2010.07.001}{\doi{10.1016/j.physrep.2010.07.001}},
\href{http://www.arXiv.org/abs/0910.1785}{\texttt{arXiv:0910.1785}}.

\bibitem{Sirunyan:2018koj}
\hrefCMSnoop {}{{CMS Collaboration}, ``Combined measurements of {Higgs} boson
  couplings in proton-proton collisions at $\sqrt{s}=13$ {TeV}'',} \textit{
  Eur. Phys. J. C} \textbf{ 79} (2019) 421,
  \href{http://dx.doi.org/10.1140/epjc/s10052-019-6909-y}{\doi{10.1140/epjc/s10052-019-6909-y}},
\href{http://www.arXiv.org/abs/1809.10733}{\texttt{arXiv:1809.10733}}.

\bibitem{Dermisek:2005ar}
\hrefCMSnoop {}{R.~Dermisek and J.~F. Gunion, ``Escaping the large fine tuning
  and little hierarchy problems in the next to minimal supersymmetric model and
  $h\rightarrow aa$ decays'',} \textit{ Phys. Rev. Lett.} \textbf{ 95} (2005)
  041801,
  \href{http://dx.doi.org/10.1103/PhysRevLett.95.041801}{\doi{10.1103/PhysRevLett.95.041801}},
\href{http://www.arXiv.org/abs/hep-ph/0502105}{\texttt{arXiv:hep-ph/0502105}}.

\bibitem{Dermisek:2006wr}
\hrefCMSnoop {}{R.~Dermisek and J.~F. Gunion, ``The {NMSSM} close to the
  {R}-symmetry limit and naturalness in $h\rightarrow aa$ decays for
  $m(a)<2m(b)$'',} \textit{ Phys. Rev. D} \textbf{ 75} (2007) 075019,
  \href{http://dx.doi.org/10.1103/PhysRevD.75.075019}{\doi{10.1103/PhysRevD.75.075019}},
\href{http://www.arXiv.org/abs/hep-ph/0611142}{\texttt{arXiv:hep-ph/0611142}}.

\bibitem{Chang:2008cw}
\hrefCMSnoop {}{S.~Chang, R.~Dermisek, J.~F. Gunion, and N.~Weiner,
  ``Nonstandard {Higgs} boson decays'',} \textit{ Ann. Rev. Nucl. Part. Sci.}
  \textbf{ 58} (2008) 75,
  \href{http://dx.doi.org/10.1146/annurev.nucl.58.110707.171200}{\doi{10.1146/annurev.nucl.58.110707.171200}},
\href{http://www.arXiv.org/abs/0801.4554}{\texttt{arXiv:0801.4554}}.

\bibitem{Gunion:1989we}
J.~F. Gunion, H.~E. Haber, G.~L. Kane, and S.~Dawson, ``The {Higgs} Hunter's
  Guide'', volume~80 of \textit{ Frontiers in Physics}.
\newblock Perseus Books, 2000.

\bibitem{King:2012tr}
\hrefCMSnoop {}{S.~F. King, M.~Muhlleitner, R.~Nevzorov, and K.~Walz, ``Natural
  {NMSSM} {Higgs} bosons'',} \textit{ Nucl. Phys. B} \textbf{ 870} (2013) 323,
  \href{http://dx.doi.org/10.1016/j.nuclphysb.2013.01.020}{\doi{10.1016/j.nuclphysb.2013.01.020}},
\href{http://www.arXiv.org/abs/1211.5074}{\texttt{arXiv:1211.5074}}.

\bibitem{Celis:2013rcs}
\hrefCMSnoop {}{A.~Celis, V.~Ilisie, and A.~Pich, ``{LHC} constraints on
  two-{Higgs} doublet models'',} \textit{ JHEP} \textbf{ 07} (2013) 053,
  \href{http://dx.doi.org/10.1007/JHEP07(2013)053}{\doi{10.1007/JHEP07(2013)053}},
\href{http://www.arXiv.org/abs/1302.4022}{\texttt{arXiv:1302.4022}}.

\bibitem{Grinstein:2013npa}
\hrefCMSnoop {}{B.~Grinstein and P.~Uttayarat, ``Carving out parameter space in
  {Type-II} two {Higgs} doublets model'',} \textit{ JHEP} \textbf{ 06} (2013)
  094,
  \href{http://dx.doi.org/10.1007/JHEP06(2013)094}{\doi{10.1007/JHEP06(2013)094}},
\href{http://www.arXiv.org/abs/1304.0028}{\texttt{arXiv:1304.0028}}.

\bibitem{Coleppa:2013dya}
\hrefCMSnoop {}{B.~Coleppa, F.~Kling, and S.~Su, ``Constraining {Type II}
  {2HDM} in light of {LHC} {Higgs} searches'',} \textit{ JHEP} \textbf{ 01}
  (2014) 161,
  \href{http://dx.doi.org/10.1007/JHEP01(2014)161}{\doi{10.1007/JHEP01(2014)161}},
\href{http://www.arXiv.org/abs/1305.0002}{\texttt{arXiv:1305.0002}}.

\bibitem{Chen:2013rba}
\hrefCMSnoop {}{C.-Y. Chen, S.~Dawson, and M.~Sher, ``Heavy {Higgs} searches
  and constraints on two {Higgs} doublet models'',} \textit{ Phys. Rev. D}
  \textbf{ 88} (2013) 015018,
  \href{http://dx.doi.org/10.1103/PhysRevD.88.015018}{\doi{10.1103/PhysRevD.88.015018}},
\href{http://www.arXiv.org/abs/1305.1624}{\texttt{arXiv:1305.1624}}.

\bibitem{Craig:2013hca}
\hrefCMSnoop {}{N.~Craig, J.~Galloway, and S.~Thomas, ``Searching for signs of
  the second {Higgs} doublet'',} (2013).
\href{http://www.arXiv.org/abs/1305.2424}{\texttt{arXiv:1305.2424}}.

\bibitem{Wang:2013sha}
\hrefCMSnoop {}{L.~Wang and X.-F. Han, ``Status of the aligned
  two-{Higgs}-doublet model confronted with the {Higgs} data'',} \textit{ JHEP}
  \textbf{ 04} (2014) 128,
  \href{http://dx.doi.org/10.1007/JHEP04(2014)128}{\doi{10.1007/JHEP04(2014)128}},
\href{http://www.arXiv.org/abs/1312.4759}{\texttt{arXiv:1312.4759}}.

\bibitem{Cao:2013gba}
J.~Cao\hrefCMSnoop {}{ {et~al.}, ``A light {Higgs} scalar in the {NMSSM}
  confronted with the latest {LHC} {Higgs} data'',} \textit{ JHEP} \textbf{ 11}
  (2013) 018,
  \href{http://dx.doi.org/10.1007/JHEP11(2013)018}{\doi{10.1007/JHEP11(2013)018}},
\href{http://www.arXiv.org/abs/1309.4939}{\texttt{arXiv:1309.4939}}.

\bibitem{Christensen:2013dra}
\hrefCMSnoop {}{N.~D. Christensen, T.~Han, Z.~Liu, and S.~Su, ``Low-mass
  {Higgs} bosons in the {NMSSM} and their {LHC} implications'',} \textit{ JHEP}
  \textbf{ 08} (2013) 019,
  \href{http://dx.doi.org/10.1007/JHEP08(2013)019}{\doi{10.1007/JHEP08(2013)019}},
\href{http://www.arXiv.org/abs/1303.2113}{\texttt{arXiv:1303.2113}}.

\bibitem{Cerdeno:2013cz}
\hrefCMSnoop {}{D.~G. Cerdeno, P.~Ghosh, and C.~B. Park, ``Probing the two
  light {Higgs} scenario in the {NMSSM} with a low-mass pseudoscalar'',}
  \textit{ JHEP} \textbf{ 06} (2013) 031,
  \href{http://dx.doi.org/10.1007/JHEP06(2013)031}{\doi{10.1007/JHEP06(2013)031}},
\href{http://www.arXiv.org/abs/1301.1325}{\texttt{arXiv:1301.1325}}.

\bibitem{Chalons:2012qe}
\hrefCMSnoop {}{G.~Chalons and F.~Domingo, ``Analysis of the {Higgs} potentials
  for two doublets and a singlet'',} \textit{ Phys. Rev. D} \textbf{ 86} (2012)
  115024,
  \href{http://dx.doi.org/10.1103/PhysRevD.86.115024}{\doi{10.1103/PhysRevD.86.115024}},
\href{http://www.arXiv.org/abs/1209.6235}{\texttt{arXiv:1209.6235}}.

\bibitem{Ahriche:2013vqa}
\hrefCMSnoop {}{A.~Ahriche, A.~Arhrib, and S.~Nasri, ``{Higgs} phenomenology in
  the two-singlet model'',} \textit{ JHEP} \textbf{ 02} (2014) 042,
  \href{http://dx.doi.org/10.1007/JHEP02(2014)042}{\doi{10.1007/JHEP02(2014)042}},
\href{http://www.arXiv.org/abs/1309.5615}{\texttt{arXiv:1309.5615}}.

\bibitem{Baglio:2014nea}
\hrefCMSnoop {}{J.~Baglio, O.~Eberhardt, U.~Nierste, and M.~Wiebusch,
  ``Benchmarks for {Higgs} pair production and heavy {Higgs} boson searches in
  the two-{Higgs}-doublet model of {Type II}'',} \textit{ Phys. Rev. D}
  \textbf{ 90} (2014) 015008,
  \href{http://dx.doi.org/10.1103/PhysRevD.90.015008}{\doi{10.1103/PhysRevD.90.015008}},
\href{http://www.arXiv.org/abs/1403.1264}{\texttt{arXiv:1403.1264}}.

\bibitem{Dumont:2014wha}
\hrefCMSnoop {}{B.~Dumont, J.~F. Gunion, Y.~Jiang, and S.~Kraml, ``Constraints
  on and future prospects for two-{Higgs}-doublet models in light of the {LHC}
  {Higgs} signal'',} \textit{ Phys. Rev. D} \textbf{ 90} (2014) 035021,
  \href{http://dx.doi.org/10.1103/PhysRevD.90.035021}{\doi{10.1103/PhysRevD.90.035021}},
\href{http://www.arXiv.org/abs/1405.3584}{\texttt{arXiv:1405.3584}}.

\bibitem{Bernon:2015qea}
J.~Bernon\hrefCMSnoop {}{ {et~al.}, ``Scrutinizing the alignment limit in
  two-{Higgs}-doublet models: m$_h$ = 125 {GeV}'',} \textit{ Phys. Rev. D}
  \textbf{ 92} (2015) 075004,
  \href{http://dx.doi.org/10.1103/PhysRevD.92.075004}{\doi{10.1103/PhysRevD.92.075004}},
\href{http://www.arXiv.org/abs/1507.00933}{\texttt{arXiv:1507.00933}}.

\bibitem{Buskulic:1993gi}
\hrefCMSnoop {}{{ALEPH} Collaboration, ``Search for a nonminimal {Higgs} boson
  produced in the reaction $e^{+} e^{-} \to$ h ${Z}^{*}$'',} \textit{ Phys.
  Lett. B} \textbf{ 313} (1993) 312,
  \href{http://dx.doi.org/10.1016/0370-2693(93)91228-F}{\doi{10.1016/0370-2693(93)91228-F}}.

\bibitem{Acciarri:1996um}
\hrefCMSnoop {}{{L3} Collaboration, ``Search for neutral {Higgs} boson
  production through the process $e^{+} e^{-} \to {Z}^{*}$ {H0}'',} \textit{
  Phys. Lett. B} \textbf{ 385} (1996) 454,
  \href{http://dx.doi.org/10.1016/0370-2693(96)00987-2}{\doi{10.1016/0370-2693(96)00987-2}}.

\bibitem{Abbiendi:2002qp}
\hrefCMSnoop {}{{OPAL} Collaboration, ``Decay mode independent searches for new
  scalar bosons with the {OPAL} detector at {LEP}'',} \textit{ Eur. Phys. J. C}
  \textbf{ 27} (2003) 311,
  \href{http://dx.doi.org/10.1140/epjc/s2002-01115-1}{\doi{10.1140/epjc/s2002-01115-1}},
  \href{http://www.arXiv.org/abs/hep-ex/0206022}{\texttt{arXiv:hep-ex/0206022}}.

\bibitem{King:2014xwa}
\hrefCMSnoop {}{S.~F. King, M.~Muhlleitner, R.~Nevzorov, and K.~Walz,
  ``Discovery prospects for {NMSSM} {Higgs} bosons at the high-energy large
  hadron collider'',} \textit{ Phys. Rev. D} \textbf{ 90} (2014) 095014,
  \href{http://dx.doi.org/10.1103/PhysRevD.90.095014}{\doi{10.1103/PhysRevD.90.095014}},
\href{http://www.arXiv.org/abs/1408.1120}{\texttt{arXiv:1408.1120}}.

\bibitem{Schael:2010aw}
\hrefCMSnoop {}{{ALEPH} Collaboration, ``Search for neutral {Higgs} bosons
  decaying into four taus at {LEP2}'',} \textit{ JHEP} \textbf{ 05} (2010) 049,
  \href{http://dx.doi.org/10.1007/JHEP05(2010)049}{\doi{10.1007/JHEP05(2010)049}},
  \href{http://www.arXiv.org/abs/1003.0705}{\texttt{arXiv:1003.0705}}.

\bibitem{Chatrchyan:2012am}
\hrefCMSnoop {}{{CMS Collaboration}, ``Search for a light pseudoscalar {Higgs}
  boson in the dimuon decay channel in pp collisions at $\sqrt{s}=7$ {TeV}'',}
  \textit{ Phys. Rev. Lett.} \textbf{ 109} (2012) 121801,
  \href{http://dx.doi.org/10.1103/PhysRevLett.109.121801}{\doi{10.1103/PhysRevLett.109.121801}},
\href{http://www.arXiv.org/abs/1206.6326}{\texttt{arXiv:1206.6326}}.

\bibitem{Dermisek:2009fd}
\hrefCMSnoop {}{R.~Dermisek and J.~F. Gunion, ``Direct production of a light
  {CP}-odd {Higgs} boson at the {Tevatron} and {LHC}'',} \textit{ Phys. Rev. D}
  \textbf{ 81} (2010) 055001,
  \href{http://dx.doi.org/10.1103/PhysRevD.81.055001}{\doi{10.1103/PhysRevD.81.055001}},
\href{http://www.arXiv.org/abs/0911.2460}{\texttt{arXiv:0911.2460}}.

\bibitem{Aaij:2018xpt}
\hrefCMSnoop {}{{LHCb Collaboration}, ``Search for a dimuon resonance in the
  $\upsilon$ mass region'',} \textit{ JHEP} \textbf{ 09} (2018) 147,
  \href{http://dx.doi.org/10.1007/JHEP09(2018)147}{\doi{10.1007/JHEP09(2018)147}},
\href{http://www.arXiv.org/abs/1805.09820}{\texttt{arXiv:1805.09820}}.

\bibitem{Sirunyan:2018mgs}
\hrefCMSnoop {}{{CMS Collaboration}, ``A search for pair production of new
  light bosons decaying into muons in proton-proton collisions at 13 {TeV}'',}
  \textit{ Phys. Lett. B} \textbf{ 796} (2019) 131,
  \href{http://dx.doi.org/10.1016/j.physletb.2019.07.013}{\doi{10.1016/j.physletb.2019.07.013}},
\href{http://www.arXiv.org/abs/1812.00380}{\texttt{arXiv:1812.00380}}.

\bibitem{Aaboud:2018fvk}
\hrefCMSnoop {}{{ATLAS Collaboration}, ``Search for {Higgs} boson decays to
  beyond-the-standard-model light bosons in four-lepton events with the {ATLAS}
  detector at $\sqrt{s}=13$ {TeV}'',} \textit{ JHEP} \textbf{ 06} (2018) 166,
  \href{http://dx.doi.org/10.1007/JHEP06(2018)166}{\doi{10.1007/JHEP06(2018)166}},
\href{http://www.arXiv.org/abs/1802.03388}{\texttt{arXiv:1802.03388}}.

\bibitem{Khachatryan:2017mnf}
\hrefCMSnoop {}{{CMS Collaboration}, ``Search for light bosons in decays of the
  125 {GeV} {Higgs} boson in proton-proton collisions at $ \sqrt{s}=8 $
  {TeV}'',} \textit{ JHEP} \textbf{ 10} (2017) 076,
  \href{http://dx.doi.org/10.1007/JHEP10(2017)076}{\doi{10.1007/JHEP10(2017)076}},
\href{http://www.arXiv.org/abs/1701.02032}{\texttt{arXiv:1701.02032}}.

\bibitem{Sirunyan:2018mbx}
\hrefCMSnoop {}{{CMS Collaboration}, ``Search for an exotic decay of the
  {Higgs} boson to a pair of light pseudoscalars in the final state of two
  muons and two $\tau$ leptons in proton-proton collisions at $ \sqrt{s}=13 $
  {TeV}'',} \textit{ JHEP} \textbf{ 11} (2018) 018,
  \href{http://dx.doi.org/10.1007/JHEP11(2018)018}{\doi{10.1007/JHEP11(2018)018}},
\href{http://www.arXiv.org/abs/1805.04865}{\texttt{arXiv:1805.04865}}.

\bibitem{Sirunyan:2018pzn}
\hrefCMSnoop {}{{CMS Collaboration}, ``Search for an exotic decay of the
  {Higgs} boson to a pair of light pseudoscalars in the final state with two b
  quarks and two $\tau$ leptons in proton-proton collisions at $\sqrt{s}=$ 13
  {TeV}'',} \textit{ Phys. Lett. B} \textbf{ 785} (2018) 462,
  \href{http://dx.doi.org/10.1016/j.physletb.2018.08.057}{\doi{10.1016/j.physletb.2018.08.057}},
\href{http://www.arXiv.org/abs/1805.10191}{\texttt{arXiv:1805.10191}}.

\bibitem{Sirunyan:2019gou}
\hrefCMSnoop {}{{CMS Collaboration}, ``Search for light pseudoscalar boson
  pairs produced from decays of the 125 {GeV} {Higgs} boson in final states
  with two muons and two nearby tracks in pp collisions at $\sqrt{s}=$ 13
  {TeV}'',} \textit{ Phys. Lett. B} \textbf{ 800} (2020) 135087,
  \href{http://dx.doi.org/10.1016/j.physletb.2019.135087}{\doi{10.1016/j.physletb.2019.135087}},
\href{http://www.arXiv.org/abs/1907.07235}{\texttt{arXiv:1907.07235}}.

\bibitem{Aad:2015oqa}
\hrefCMSnoop {}{{ATLAS Collaboration}, ``Search for {Higgs} bosons decaying to
  $aa$ in the $\mu\mu\tau\tau$ final state in pp collisions at $\sqrt{s} = $ 8
  {TeV} with the {ATLAS} experiment'',} \textit{ Phys. Rev. D} \textbf{ 92}
  (2015) 052002,
  \href{http://dx.doi.org/10.1103/PhysRevD.92.052002}{\doi{10.1103/PhysRevD.92.052002}},
\href{http://www.arXiv.org/abs/1505.01609}{\texttt{arXiv:1505.01609}}.

\bibitem{Aaboud:2018esj}
\hrefCMSnoop {}{{ATLAS Collaboration}, ``Search for {Higgs} boson decays into a
  pair of light bosons in the $bb\mu\mu$ final state in pp collision at
  $\sqrt{s} = $13 {TeV} with the {ATLAS} detector'',} \textit{ Phys. Lett. B}
  \textbf{ 790} (2019) 1,
  \href{http://dx.doi.org/10.1016/j.physletb.2018.10.073}{\doi{10.1016/j.physletb.2018.10.073}},
\href{http://www.arXiv.org/abs/1807.00539}{\texttt{arXiv:1807.00539}}.

\bibitem{Aaboud:2018iil}
\hrefCMSnoop {}{{ATLAS Collaboration}, ``Search for the {Higgs} boson produced
  in association with a vector boson and decaying into two spin-zero particles
  in the $h \rightarrow aa \rightarrow 4b$ channel in pp collisions at
  $\sqrt{s} = 13$ {TeV} with the {ATLAS} detector'',} \textit{ JHEP} \textbf{
  10} (2018) 031,
  \href{http://dx.doi.org/10.1007/JHEP10(2018)031}{\doi{10.1007/JHEP10(2018)031}},
\href{http://www.arXiv.org/abs/1806.07355}{\texttt{arXiv:1806.07355}}.

\bibitem{deFlorian:2016spz}
\hrefCMSnoop {}{{LHC Higgs Cross Section Working Group}, ``Handbook of {LHC
  Higgs} cross sections: 4. deciphering the nature of the {Higgs} sector'',}
  \textit{ CERN} (2016)
  \href{http://dx.doi.org/10.23731/CYRM-2017-002}{\doi{10.23731/CYRM-2017-002}},
\href{http://www.arXiv.org/abs/1610.07922}{\texttt{arXiv:1610.07922}}.

\bibitem{Bernon:2014nxa}
\hrefCMSnoop {}{J.~Bernon, J.~F. Gunion, Y.~Jiang, and S.~Kraml, ``Light
  {Higgs} bosons in two-{Higgs}-doublet models'',} \textit{ Phys. Rev. D}
  \textbf{ 91} (2015) 075019,
  \href{http://dx.doi.org/10.1103/PhysRevD.91.075019}{\doi{10.1103/PhysRevD.91.075019}},
\href{http://www.arXiv.org/abs/1412.3385}{\texttt{arXiv:1412.3385}}.

\bibitem{Khachatryan:2016bia}
\hrefCMSnoop {}{{CMS Collaboration}, ``The {CMS} trigger system'',} \textit{
  JINST} \textbf{ 12} (2017) P01020,
  \href{http://dx.doi.org/10.1088/1748-0221/12/01/P01020}{\doi{10.1088/1748-0221/12/01/P01020}},
\href{http://www.arXiv.org/abs/1609.02366}{\texttt{arXiv:1609.02366}}.

\bibitem{Chatrchyan:2008zzk}
\hrefCMSnoop {}{{CMS Collaboration}, ``The {CMS} experiment at the {CERN}
  {LHC}'',} \textit{ JINST} \textbf{ 3} (2008) S08004,
\href{http://dx.doi.org/10.1088/1748-0221/3/08/S08004}{\doi{10.1088/1748-0221/3/08/S08004}}.

\bibitem{mg_amcnlo}
J.~Alwall\hrefCMSnoop {}{ {et~al.}, ``The automated computation of tree-level
  and next-to-leading order differential cross sections, and their matching to
  parton shower simulations'',} \textit{ JHEP} \textbf{ 07} (2014) 079,
  \href{http://dx.doi.org/10.1007/JHEP07(2014)079}{\doi{10.1007/JHEP07(2014)079}},
\href{http://www.arXiv.org/abs/1405.0301}{\texttt{arXiv:1405.0301}}.

\bibitem{Sjostrand:2014zea}
T.~Sj{\"o}strand\hrefCMSnoop {}{ {et~al.}, ``An introduction to {PYTHIA
  8.2}'',} \textit{ Comput. Phys. Commun.} \textbf{ 191} (2015) 159,
  \href{http://dx.doi.org/10.1016/j.cpc.2015.01.024}{\doi{10.1016/j.cpc.2015.01.024}},
\href{http://www.arXiv.org/abs/1410.3012}{\texttt{arXiv:1410.3012}}.

\bibitem{Khachatryan:2015pea}
\hrefCMSnoop {}{{CMS Collaboration}, ``Event generator tunes obtained from
  underlying event and multiparton scattering measurements'',} \textit{ Eur.
  Phys. J. C} \textbf{ 76} (2016) 155,
  \href{http://dx.doi.org/10.1140/epjc/s10052-016-3988-x}{\doi{10.1140/epjc/s10052-016-3988-x}},
\href{http://www.arXiv.org/abs/1512.00815}{\texttt{arXiv:1512.00815}}.

\bibitem{Ball:2017nwa}
\hrefCMSnoop {}{{NNPDF} Collaboration, ``Parton distributions from
  high-precision collider data'',} \textit{ Eur. Phys. J. C} \textbf{ 77}
  (2017) 663,
  \href{http://dx.doi.org/10.1140/epjc/s10052-017-5199-5}{\doi{10.1140/epjc/s10052-017-5199-5}},
\href{http://www.arXiv.org/abs/1706.00428}{\texttt{arXiv:1706.00428}}.

\bibitem{GEANT}
\hrefCMSnoop {}{{GEANT4} Collaboration, ``{\GEANTfour}---a simulation
  toolkit'',} \textit{ Nucl. Instrum. Meth. A} \textbf{ 506} (2003) 250,
\href{http://dx.doi.org/10.1016/S0168-9002(03)01368-8}{\doi{10.1016/S0168-9002(03)01368-8}}.

\bibitem{Sirunyan:2017ulk}
\hrefCMSnoop {}{{CMS Collaboration}, ``Particle-flow reconstruction and global
  event description with the {CMS} detector'',} \textit{ JINST} \textbf{ 12}
  (2017) P10003,
  \href{http://dx.doi.org/10.1088/1748-0221/12/10/P10003}{\doi{10.1088/1748-0221/12/10/P10003}},
\href{http://www.arXiv.org/abs/1706.04965}{\texttt{arXiv:1706.04965}}.

\bibitem{Cacciari:2008gp}
\hrefCMSnoop {}{M.~Cacciari, G.~P. Salam, and G.~Soyez, ``The anti-\kt jet
  clustering algorithm'',} \textit{ JHEP} \textbf{ 04} (2008) 063,
  \href{http://dx.doi.org/10.1088/1126-6708/2008/04/063}{\doi{10.1088/1126-6708/2008/04/063}},
  \href{http://www.arXiv.org/abs/0802.1189}{\texttt{arXiv:0802.1189}}.

\bibitem{Cacciari:2011ma}
\hrefCMSnoop {}{M.~Cacciari, G.~P. Salam, and G.~Soyez, ``{FastJet} user
  manual'',} \textit{ Eur. Phys. J. C} \textbf{ 72} (2012) 1896,
  \href{http://dx.doi.org/10.1140/epjc/s10052-012-1896-2}{\doi{10.1140/epjc/s10052-012-1896-2}},
\href{http://www.arXiv.org/abs/1111.6097}{\texttt{arXiv:1111.6097}}.

\bibitem{Sirunyan:2018fpa}
\hrefCMSnoop {}{{CMS Collaboration}, ``Performance of the {CMS} muon detector
  and muon reconstruction with proton-proton collisions at $\sqrt{s}=$ 13
  {TeV}'',} \textit{ JINST} \textbf{ 13} (2018) P06015,
  \href{http://dx.doi.org/10.1088/1748-0221/13/06/P06015}{\doi{10.1088/1748-0221/13/06/P06015}},
\href{http://www.arXiv.org/abs/1804.04528}{\texttt{arXiv:1804.04528}}.

\bibitem{Chatrchyan:2011ds}
\hrefCMSnoop {}{{CMS Collaboration}, ``Determination of jet energy calibration
  and transverse momentum resolution in {CMS}'',} \textit{ JINST} \textbf{ 6}
  (2011) P11002,
  \href{http://dx.doi.org/10.1088/1748-0221/6/11/P11002}{\doi{10.1088/1748-0221/6/11/P11002}},
\href{http://www.arXiv.org/abs/1107.4277}{\texttt{arXiv:1107.4277}}.

\bibitem{CMS-PAS-JME-18-001}
\href {https://cds.cern.ch/record/2683784}{{CMS Collaboration}, ``Pileup
  mitigation at {CMS} in 13 {TeV} data'',} CMS Physics Analysis Summary
  CMS-PAS-JME-18-001, 2019.

\bibitem{Chatrchyan:2012jua}
\hrefCMSnoop {}{{CMS Collaboration}, ``Identification of b quark jets with the
  {CMS} experiment'',} \textit{ JINST} \textbf{ 8} (2013) P04013,
  \href{http://dx.doi.org/10.1088/1748-0221/8/04/P04013}{\doi{10.1088/1748-0221/8/04/P04013}},
\href{http://www.arXiv.org/abs/1211.4462}{\texttt{arXiv:1211.4462}}.

\bibitem{Sirunyan:2017ezt}
\hrefCMSnoop {}{{CMS Collaboration}, ``Identification of heavy-flavour jets
  with the {CMS} detector in pp collisions at 13 {TeV}'',} \textit{ JINST}
  \textbf{ 13} (2018) P05011,
  \href{http://dx.doi.org/10.1088/1748-0221/13/05/P05011}{\doi{10.1088/1748-0221/13/05/P05011}},
\href{http://www.arXiv.org/abs/1712.07158}{\texttt{arXiv:1712.07158}}.

\bibitem{Sirunyan:2018pgf}
\hrefCMSnoop {}{{CMS Collaboration}, ``Performance of reconstruction and
  identification of $\tau$ leptons decaying to hadrons and $\nu_\tau$ in pp
  collisions at $\sqrt{s}=$ 13 {TeV}'',} \textit{ JINST} \textbf{ 13} (2018)
  P10005,
  \href{http://dx.doi.org/10.1088/1748-0221/13/10/P10005}{\doi{10.1088/1748-0221/13/10/P10005}},
\href{http://www.arXiv.org/abs/1809.02816}{\texttt{arXiv:1809.02816}}.

\bibitem{CMS:2011aa}
\hrefCMSnoop {}{{CMS Collaboration}, ``Measurement of the inclusive {$\PW$} and
  {$\PZ$} production cross sections in pp collisions at $\sqrt{s}=7$ {TeV} with
  the {CMS} experiment'',} \textit{ JHEP} \textbf{ 10} (2011) 132,
  \href{http://dx.doi.org/10.1007/JHEP10(2011)132}{\doi{10.1007/JHEP10(2011)132}},
\href{http://www.arXiv.org/abs/1107.4789}{\texttt{arXiv:1107.4789}}.

\bibitem{PDG2018}
\hrefCMSnoop {}{{Particle Data Group}, M.~Tanabashi {et~al.}, ``Review of
  particle physics'',} \textit{ Phys. Rev. D} \textbf{ 98} (2018) 030001,
  \href{http://dx.doi.org/10.1103/PhysRevD.98.030001}{\doi{10.1103/PhysRevD.98.030001}}.

\bibitem{CMS-PAS-LUM-17-001}
\href {https://cds.cern.ch/record/2257069}{{{CMS}} Collaboration, ``{CMS}
  luminosity measurements for the 2016 data taking period'',} CMS Physics
  Analysis Summary CMS-PAS-LUM-17-001, 2017.

\bibitem{Sirunyan:2018nqx}
\hrefCMSnoop {}{{CMS Collaboration}, ``Measurement of the inelastic
  proton-proton cross section at $ \sqrt{s}=13$ {TeV}'',} \textit{ JHEP}
  \textbf{ 07} (2018) 161,
  \href{http://dx.doi.org/10.1007/JHEP07(2018)161}{\doi{10.1007/JHEP07(2018)161}},
\href{http://www.arXiv.org/abs/1802.02613}{\texttt{arXiv:1802.02613}}.

\bibitem{Junk:1999kv}
\hrefCMSnoop {}{T.~Junk, ``Confidence level computation for combining searches
  with small statistics'',} \textit{ Nucl. Instrum. Meth. A} \textbf{ 434}
  (1999) 435,
  \href{http://dx.doi.org/10.1016/S0168-9002(99)00498-2}{\doi{10.1016/S0168-9002(99)00498-2}},
\href{http://www.arXiv.org/abs/hep-ex/9902006}{\texttt{arXiv:hep-ex/9902006}}.

\bibitem{Read:2002hq}
\hrefCMSnoop {}{A.~L. Read, ``Presentation of search results: The
  {CL}$_{\rm{s}}$ technique'',} \textit{ J. Phys. G} \textbf{ 28} (2002) 2693,
\href{http://dx.doi.org/10.1088/0954-3899/28/10/313}{\doi{10.1088/0954-3899/28/10/313}}.

\bibitem{CMS-NOTE-2011-005}
\href {https://cds.cern.ch/record/1379837}{{ATLAS and CMS Collaborations, and
  The LHC Higgs Combination Group}, ``Procedure for the {LHC} {Higgs} boson
  search combination in {Summer} 2011'',} Technical Report CMS-NOTE-2011-005.
  ATL-PHYS-PUB-2011-11, 2011.

\bibitem{Haisch:2018kqx}
\hrefCMSnoop {}{U.~Haisch, J.~F. Kamenik, A.~Malinauskas, and M.~Spira,
  ``Collider constraints on light pseudoscalars'',} \textit{ JHEP} \textbf{ 03}
  (2018) 178,
  \href{http://dx.doi.org/10.1007/JHEP03(2018)178}{\doi{10.1007/JHEP03(2018)178}},
\href{http://www.arXiv.org/abs/1802.02156}{\texttt{arXiv:1802.02156}}.

\end{thebibliography}\endgroup
